# Deriving Dependently-Typed OOP from First Principles

Extended Version with Additional Appendices

DAVID BINDER, University of Tübingen, Germany
INGO SKUPIN, University of Tübingen, Germany
TIM SÜBERKRÜB, Aleph Alpha Research at IPAI, Germany
KLAUS OSTERMANN, University of Tübingen, Germany

The *expression problem* describes how most types can easily be extended with new ways to *produce* the type or new ways to *consume* the type, but not both. When abstract syntax trees are defined as an algebraic data type, for example, they can easily be extended with new consumers, such as *print* or *eval*, but adding a new constructor requires the modification of all existing pattern matches. The expression problem is one way to elucidate the difference between functional or data-oriented programs (easily extendable by new consumers) and object-oriented programs (easily extendable by new producers). This difference between programs which are extensible by new producers or new consumers also exists for dependently typed programming, but with one core difference: Dependently-typed programming almost exclusively follows the functional programming model and not the object-oriented model, which leaves an interesting space in the programming language landscape unexplored. In this paper, we explore the field of dependently-typed object-oriented programming by *deriving it from first principles* using the principle of duality. That is, we do not extend an existing object-oriented formalism with dependent types in an ad-hoc fashion, but instead start from a familiar data-oriented language and derive its dual fragment by the systematic use of defunctionalization and refunctionalization. Our central contribution is a dependently typed calculus which contains two dual language fragments. We provide type- and semantics-preserving transformations between these two language fragments: defunctionalization and refunctionalization. We have implemented this language and these transformations and use this implementation to explain the various ways in which constructions in dependently typed programming can be explained as special instances of the general phenomenon of duality.

## 1 INTRODUCTION

There are many programming paradigms, but dependently typed programming languages almost exclusively follow the functional programming model. In this paper, we show why dependently-typed programming languages should also include object-oriented principles, and how this can be done. One of the main reasons why object-oriented features should be included is a consequence of how the complexity of the domain is modeled in the functional and object-oriented paradigm. Functional programmers structure the domain using data types defined by their constructors, whereas object-oriented programmers structure the domain using classes and interfaces defined by methods. This choice has important implications for the extensibility properties of large programs, which are only more accentuated for dependently typed programs.

Why do most dependently typed languages follow the functional style? One of the main reasons is that dependent type theories, on which a lot of them are based, are best studied for functional programming languages. Our challenge, then, is to develop a dependently-typed object-oriented

Authors' addresses: David Binder, Department of Computer Science, University of Tübingen, Sand 14, Tübingen, 72076, Germany, david.binder@uni-tuebingen.de; Ingo Skupin, Department of Computer Science, University of Tübingen, Sand 14, Tübingen, 72076, Germany, skupin@informatik.uni-tuebingen.de; Tim Süberkrüb, Aleph Alpha Research at IPAI, Grenzhöfer Weg 36, Heidelberg, 69123, Germany, tim.sueberkrueb@aleph-alpha-ip.ai; Klaus Ostermann, Department of Computer Science, University of Tübingen, Sand 14, Tübingen, 72076, Germany, klaus.ostermann@uni-tuebingen.de.

calculus that can serve as the foundation for object-oriented dependently-typed programming languages. Instead of specifying this calculus in an ad-hoc fashion, we want to use de- and refunctionalization as systematic tools to *derive* an object-oriented language fragment from its functional counterpart. We want to show how object-oriented programming is *dual* to functional programming, that this duality extends from non-dependent programming languages to dependently typed programming languages, and that we can use this duality to derive our calculus.

### 1.1 Data and Codata: The Essence of Functional and Object-Oriented Programming

How can functional programming (FP) and object-oriented programming (OOP) be dual, if there is no precise definition of these two paradigms? We have to define what we mean by functional and object-oriented programming if we want to get a precise research question. For the purposes of this paper, and other reasonable definitions notwithstanding, we focus on the differences in program decomposition between the two paradigms.[1] For us, the essence of functional programming is programming with algebraic data types and pattern matching, whereas the essence of object-oriented programming is programming against interfaces, which correspond to the type-theoretic concept of codata and copattern matching. This definition is not novel but follows similar observations by Cook [1990, 2009] and Downen et al. [2019]. A potentially confusing but important aspect of this definition is that first-class functions are in the object-oriented space, since they are a particular form of codata (and $\lambda$ is a particular form of copattern matching). In the rest of this subsection, we elaborate on this definition.

Let us verify first that this definition captures the essence of FP. An essential part of the programming experience in statically typed functional languages like OCaml, Scala, Haskell or SML, but also proof assistants like Coq, Agda, Idris and Lean, is modeling the domain with algebraic data types. Algebraic data types consist of product types like structs and records, sum types and enums, and recursive types like lists, which together form the essential vocabulary with which programmers in those languages express themselves. The dependently typed languages in this list extend this vocabulary by allowing data types to be *indexed*; the vector type, for example, is indexed over the number of its elements.

That OOP can be identified with codata types is less obvious, so we will introduce them with a bit more detail. Data types and codata types differ in how they are defined: Whereas a data type is defined by its constructors, i.e. all the ways in which terms of that type can be constructed, a codata type is defined by all the ways it can be observed. One type which is defined by its two canonical observations is the type of infinite streams. We can either observe the head of a stream, yielding one element, or we can observe the tail, yielding a new stream. Equivalently, we can say that every stream has to implement the stream interface which requires a head and a tail method. Instead of this object-oriented terminology, we use the type-theoretic jargon and the following syntax for defining the type of streams:

```
codata Stream(a: Type) {
  Stream(a).head(a: Type): a,
  Stream(a).tail(a: Type): Stream(a) }
```

```
codef Ones: Stream(Nat) {
  head(_) => S(Z),
  tail(_) => Ones }
```

The right-hand side shows how to construct a stream by implementing the stream interface, i.e. by saying how it will behave on the head and the tail observation. This particular stream models an infinite sequence of ones. The syntactic construct we use here is called copattern matching [Abel et al. 2013] and is the precise dual of pattern matching.

[1] Such a definition necessarily reduces the differences between the two paradigms to only one aspect, but this reduction is hopefully also illuminating. Focusing on another difference, and, for example, analyzing how subtyping can influence the design of dependently typed programming language would be another interesting research question.

We mentioned that according to our definition of FP and OOP, first-class functions counter-intuitively belong to the object-oriented space, so let us substantiate that claim. Programmers in functional programming languages can define many types, but they usually cannot define the function type. Functions can be defined, however, using codata types: A function is just an object which implements an interface with one apply method. For example, functions from natural numbers to Booleans, and the constant function which always returns true are defined in the following way:

```
codata Fun { ap(x: Nat): Bool }                codef ConstTrue: Fun { ap(_) => True }
```

The research question that motivated this paper is this: If functional programming can be and has been extended to dependent functional programming, can object-oriented programming be similarly extended? Codata has been introduced to many proof assistants before, but for an entirely different purpose. The purpose was to model certain infinite structures and coinductive objects, not to program in an object-oriented style. In this paper, we are interested in this second aspect, and we are (to the best of our knowledge) the first ones to discuss this question in detail. To approach this question in a principled way, we need an additional technical tool, defunctionalization and refunctionalization, which we introduce in the next section.

## 1.2 De- and Refunctionalization: A Tool for Systematic Language Design

Now that we have introduced two alternative programming paradigms, let us look at how one paradigm can express programs in the other paradigm. One way in which object-oriented programmers have often represented the functional style is with the visitor pattern [Gamma et al. 1995]. Later, Downen et al. [2019] showed how the visitor pattern can be used as a compilation technique for data and codata types; using the visitor pattern, they can compile functional programs to object-oriented programs, and using a related tabulation technique they can compile object-oriented programs to functional ones. In this paper, we use an alternative technique: defunctionalization and refunctionalization.

Defunctionalization [Danvy and Nielsen 2001; Reynolds 1972] is a whole-program transformation which eliminates higher-order functions by replacing lambda abstractions by constructors of a data type, together with a top-level `apply` function. Refunctionalization [Danvy and Millikin 2009] is its partial inverse, and re-introduces higher-order functions by replacing occurrences of the constructors by lambda abstractions. We already observed in the previous section that the function type is just one instance of a codata type. Based on this observation, Rendel et al. [2015] showed that defunctionalization and refunctionalization can be generalized to arbitrary data and codata types, which makes these transformations both more powerful and more symmetric since refunctionalization is now a full inverse instead of a partial one.

Let us look at an example of how these generalized defunctionalization and refunctionalization transformations work. In Figure 1a we have defined Booleans as a data type with two constructors, and negation by pattern matching on `True` and `False`. For negation we use syntax familiar to object-oriented programmers: negating a boolean $b$ can be written as $b$.`neg`. Refunctionalizing this program results in the program in Figure 1b. In this representation, negation is the single observation of a codata type, and `True` and `False` are defined as objects implementing this interface.

```
data Bool { True, False }
def Bool.neg: Bool {
   True => False,
   False => True }
```

(a) Functional programming style.

```
codata Bool { neg: Bool }
codef True: Bool { neg => False }
codef False: Bool { neg => True }
```

(b) Object-oriented style.

Fig. 1. Two representations of the same program.

One way to visualize how defunctionalization and refunctionalization work is to think of each type as a matrix. The two programs of Figure 1, for example, can be represented by the following matrix:

| **Bool** | *True* | *False* |
|---|---|---|
| *neg* | False | True |

The rows of the matrix enumerate all the ways elements of the type can be consumed, whereas the columns enumerate the ways in which elements of the type can be constructed. The cells of the matrix specify the result of an interaction between one of each. Data types and codata types are then just two different linear presentations of this type-matrix, and defunctionalization and refunctionalization transpose the linearization.

In this paper, we use defunctionalization and refunctionalization not as a compilation technique, but as a *tool for systematic language design*. These transformations are only total in a language where the data and codata fragments of the language are equally expressive. We can therefore use them to systematically *derive* the codata fragment of an object-oriented dependently-typed language by starting from a familiar design for dependent data types and pattern matching, and refunctionalizing programs in that language.

### 1.3 A Minimal Dependently-Typed Example

Let us now extend the example from the previous section by a simple proof that negation is an involution, i.e. that applying negation twice is the identity. We look at this example first from the familiar point of view of functional programming, and then from the more unfamiliar point of view of dependently-typed object-oriented programming. These two dual presentations are not artificially constructed but inter-derived using de- and refunctionalization introduced in the previous section. In the accompanying implementation that we provide, each version can be automatically transformed into the other presentation at the click of a button.

In the functional decomposition, shown in Figure 2a, we use the Martin-Löf equality type Eq($a$ : Type, $x$ $y$ : $a$) to express propositional equality. The way we defined the proposition that negating a boolean twice is the identity function is interesting. Instead of a dependent function, it is formulated more directly as an elimination on a named boolean `self` which yields a proof that `self` is equal to `self` twice-negated, i.e. `self.neg.neg`. The proof pattern matches on `True` and `False` and returns the `Refl` constructor in each branch.

```
data Eq(a: Type, x y: a) {
   Refl(a: Type, x: a): Eq(a, x, x) }
data Bool { True, False }
def Bool.neg: Bool {
   True => False,
   False => True }
def (self: Bool).neg_inverse
   : Eq(Bool, self, self.neg.neg) {
   True => Refl(Bool, True),
   False => Refl(Bool, False) }
```

(a) Functional programming style.

```
data Eq(a: Type, x y: a) {
   Refl(a: Type, x: a): Eq(a, x, x) }
codata Bool {
   neg: Bool,
   (self: Bool).neg_inverse
     : Eq(Bool, self, self.neg.neg) }
codef True: Bool {
   neg => False,
   neg_inverse => Refl(Bool, True) }
codef False: Bool {
   neg => True,
   neg_inverse => Refl(Bool, False) }
```

(b) Object-oriented style.

Fig. 2. Extending Figure 1 with proofs.

In the object-oriented decomposition, shown in Figure 2b, we have kept the definition of the Martin-Löf equality type. The definition of Booleans, on the other hand, has changed dramatically. Booleans are now defined via the two observations that we defined in the original program: negation and the proof that negating a boolean twice is the identity. Instead of two canonical constructors `True` and `False` we now have two mutually recursive top-level definitions of `True` and `False`. This means that we are now free to add new Booleans without changing the definition of the type

Bool, thus we could just add another object and implement the negation operation together with a proof of its correctness, i.e. a proof that applying it twice yields the original element.

Defining objects by interfaces that they have to implement is of course familiar. However, also including proofs of correctness in those interfaces, and looking at familiar types like Booleans in this flipped representation is novel. In this article, we invite you to follow us on an exploration of the duality of these two programming styles and to discover both the expressive power we get and the sometimes subtle problems we encounter and the restrictions we have to impose.

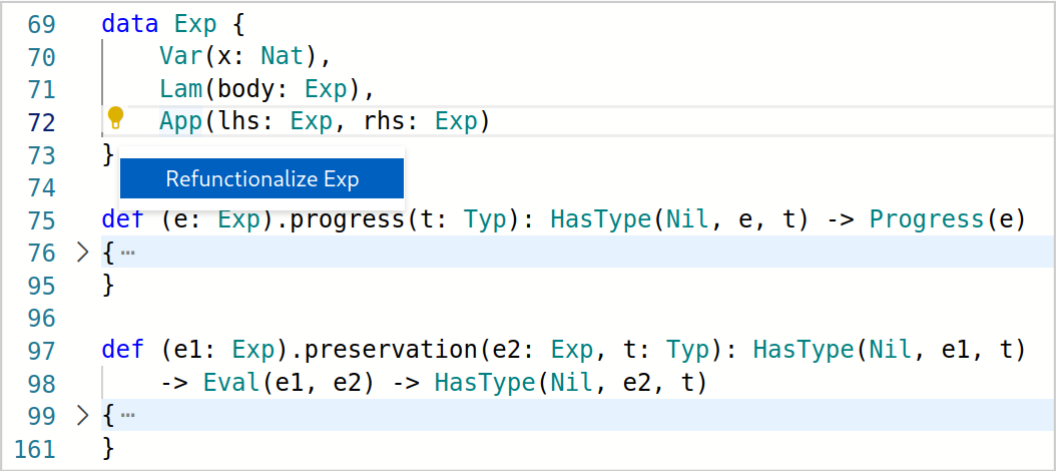

Fig. 3. Screenshot of the online IDE available at polarity-lang.github.io/oopsla24.

## 1.4 Overview

The remainder of this article is structured as follows:

- Building on our minimal example, we present dependently typed object-oriented programming in Section 2. Our language consists of two fragments, a functional/data-oriented fragment and an object-oriented/codata fragment, and the specification of these two fragments is dictated by the requirement that defunctionalization and refunctionalization are total and semantic-preserving transformations.
- In Section 3 we evaluate the expressive power and the extensibility properties of our system using a case study of a dependently typed web server. In Appendix C, we evaluate our design and implementation by a formalization of type preservation for a simple expression language respectively full type soundness of the simply typed lambda calculus. Since we have an available online implementation, the reader can choose to flip any of the involved types from the data to the codata representation, and vice-versa, and observe the resulting program.
- In Section 4 we discuss the constraints on the design of the type system that we had to observe because we want our system to be closed under de- and refunctionalization. For each such constraint, we discuss both the problem and the solution that we have chosen for our formalization.
- In Sections 5 and 6 we present all the formal details. We specify a declarative formalization in the style of Martin-Löf in Section 5 and the details of the defunctionalization and refunctionalization algorithms in Section 6.
- We discuss future work in Section 7, related work in Section 8 and conclude in Section 9.

We have implemented the language, and the defunctionalization and refunctionalization algorithms presented in this paper. We make an IDE available in the browser (cf. Figure 3) which supports the defunctionalization and refunctionalization transformations as code actions using the language server protocol (LSP).

## 2 DEPENDENTLY-TYPED OBJECT-ORIENTED PROGRAMMING

Typical programs written in object-oriented and functional languages have many differences. We will now look at how these differences appear when we consider programs written in the object-oriented style.

### 2.1 Method Call Syntax and Self Parameters

Object-oriented programmers are familiar with the *method call syntax* $o.\texttt{f}(e)$ to invoke a method `f` taking an argument $e$ on an object $o$. This is sometimes presented under the name "uniform function call syntax" as an alternative notation for the call $\texttt{f}(o, e)$, where the first argument of the method `f` is called the *self parameter*. We take this simple syntactic observation, see how we have to modify it to the dependently typed setting, and how this influenced our design of codata types.

As a starting point, we take the Agda proof from Figure 4. In that proof, the Booleans are defined as a datatype, negation is defined as a non-dependent function from Booleans to Booleans, and `neg_inverse` is defined as a dependent function from a boolean $x$ to a proof that $x$ is equal to $\texttt{neg}(\texttt{neg}\ x)$. In our system, we want to express both `neg` and `neg_inverse` without using non-dependent or dependent functions. For the non-dependent case, we can express negation directly as an observation on Booleans:

```
import Agda.Builtin.Equality
open Agda.Builtin.Equality

data Bool : Set where
  true : Bool
  false : Bool

neg : Bool -> Bool
neg true = false
neg false = true

neg_inverse
  : (x : Bool) -> x ≡ neg (neg x)
neg_inverse true = refl
neg_inverse false = refl
```

Fig. 4. Proving that negation is an involution in Agda.

```
def Bool.neg: Bool {
   True => False,
   False => True }
```

If we want to express the dependent function `neg_inverse` as an observation on Booleans in a similar way, then we have to add a feature, *self parameters*. We can bind the term that we observe to a variable, which we have here called *self*, and use this variable in the return type of the observation:

```
def (self: Bool).neg_inverse: Eq(Bool, self, self.neg.neg) {
   True => Refl(Bool, True),
   False => Refl(Bool, False) }
```

The refunctionalization of these methods with self-parameters dictates the first feature of dependent codata types: self-parameters in destructors.

```
codata Bool { neg: Bool, (self: Bool).neg_inverse: Eq(Bool, self, self.neg.neg) }
```

It is self-parameters in codata declarations which give us the expressive power to properly represent verified interfaces and classes in an object-oriented style.

### 2.2 Verified Interfaces and Classes

When verifying data structures and algorithms in dependently typed languages, we can choose between two general approaches: intrinsic and extrinsic verification. Using intrinsic verification, we define the data structure or algorithm together with its correctness proofs. To intrinsically verify data structures, we commonly express properties using type indices, such as the number of elements contained in a length-indexed list. We can employ a similar approach in an object-oriented style using indexed codata types [Thibodeau et al. 2016]. Using type indices, we can ensure that observations are called only on objects which are in the right state. This can be seen in the following example, where the observation `read` may be called only on non-empty buffers:[2]

[2]Note that the parameter $n$ occurs bound in the type `Buffer(S(n))` on which we can call the observation `read`, and is bound in the argument list `read(n: Nat)`.

```
class spec: PR
    public methods:
      store: X × A ⟶ X
      read: X ⟶ {error} + A
      empty: X ⟶ X
    assertions:
      s.empty.read = error
      s.read = error
         ⊢ s.store(a).read = a
      s.read = a
         ⊢ s.store(b).read = a
    creation:
      new.read = error
  end class spec
```

(a) Original specification

```
codata PR {
   store(a: A): PR,
   read: MaybeA,
   empty: PR,
   -- | Reading from the empty buffer yields an error
   (s: PR).assert_empty: Eq(MaybeA, s.empty.read, Error),
   -- | We can store an element into an empty buffer
   (s: PR).assert_empty_store(a: A)
      : Eq(MaybeA, s.read, Error) -> Eq(MaybeA, s.store(a).read, Just(a)),
   -- | We cannot replace the element in the buffer without calling `empty`
   (s: PR).assert_persistent(a b: A)
      : Eq(MaybeA, s.read, Just(a)) -> Eq(MaybeA, s.store(b).read, Just(a)) }
```

(b) Implementation in our system.

Fig. 5. Persistent read (PR) specification for one-element buffers from Jacobs [1995].

```
codata Buffer(m: Nat) {
   Buffer(S(n)).read(n: Nat): Pair(Bool, Buffer(n)) }
codef EmptyBuffer: Buffer(Z) { read(n) absurd }
codef Singleton(b: Bool): Buffer(S(Z)) { read(n) => MkPair(Bool, Buffer(Z), b, EmptyBuffer) }
```

We can see that, as usual for dependent (co)pattern matching, infeasible pattern matches may arise which need to be marked as `absurd`. That is, when we implement the buffer interface for the empty buffer we don't have to implement the `read` method since it can never be called.

However, in this work, we go beyond indexed codata types, which admit an intrinsic verification style. We also want to support the extrinsic approach, where we want to separate our objects from their specifications. Jacobs [1995] provides us with an initial concept of how to attain that goal. He proposes a system of coalgebraic specifications that can be used to verify object-oriented classes. As an example, Figure 5a shows a coalgebraic specification for a one-element buffer that exhibits *persistent read* (PR) behavior: After an element has been stored, it cannot be replaced using the `store` method. Instead, one needs to call the method `empty` to explicitly empty the buffer. Reading from the empty buffer returns an error. This specification of the buffer is given as a set of assertions that reference the buffer state `s`.

In our system, we can realize this concept using self-parameters on destructors, allowing us to express specifications as observations on codata types. The codata type in Figure 5 defines the verified interface for persistent read buffers in our system. Similarly, we can apply this approach to express verified interfaces such as functors or monads.

## 2.3 Dependent Functions

Unlike in most other dependent type theories, the Π-type of dependent functions is not part of our core theory, but can be defined in a library. The Π-type is defined as a codata type indexed over a type family `p`, for which we use the ordinary non-dependent function type:

```
-- | Non-dependent Functions
codata Fun(a b: Type) {
   Fun(a, b).ap(a b: Type, x: a): b }
-- | Dependent Functions
codata Π(a: Type, p: a -> Type) {
   Π(a, p).dap(a: Type, p: a -> Type, x: a): p.ap(a, Type, x) }
```

We propose that both dependent and non-dependent functions should be user-defined instead of built-in. The designers of Java decided to follow this approach when they introduced lambda abstractions as instances of *functional interfaces* [Goetz et al. 2014; Setzer 2003] in Java 8. This shows that our proposal is not radical, and we think it is also useful. Apart from reducing the complexity of the core language, they simplify the situation if we have more than one function type. This is the case in substructural systems where we have linear and non-linear functions. For instance, in the Rust programming language, there are three different built-in function traits `Fn`, `FnOnce`, and `FnMut`, which differ in the modality of the receiver[3].

### 2.4 Weak and Strong Dependent Pairs

In the previous section, we showed how to define the Π-type. For the Π-type we had no choice but to define it as a codata type[4], but for the Σ-type we can choose whether to model it as a data or codata type. This choice distinguishes weak and strong Σ-types [Howard 1980]: *Strong* Σ-types are defined as a codata type with two projections, where the second projection mentions the result of the first projection in its return type; *weak* Σ-types, by contrast, are defined as a data type with one constructor which pairs the first and second element. This difference is more obvious if we first consider the case of non-dependent pairs, which can also be written as either a data or codata type.

```
data ×₊(A B: Type) {
   Pair(A B: Type, x: A, y: B): ×₊(A, B) }
def ×₊(A, B).π₁(A B: Type): A {
   Pair(_, _, x, y) => x }
def ×₊(A, B).π₂(A B: Type): B {
   Pair(_, _, x, y) => y }
```

```
codata ×₋(A B: Type) {
   ×₋(A, B).π₁(A B: Type): A,
   ×₋(A, B).π₂(A B: Type): B }
codef Pair(A B: Type, x: A, y: B): ×₋(A, B) {
   π₁(_, _) => x,
   π₂(_, _) => y }
```

These two representations can be obtained from each other by defunctionalization and refunctionalization. This is still the case when we generalize non-dependent pairs to the Σ-type. Similar to the Π-type in Section 2.3, the Σ-type is indexed by a type family $T$. As a data type, it is defined by one constructor `Pair` which takes the type family $T$, an element $x$ of type $A$ and a witness $w$ as arguments. As a codata type, we still have two projections $\pi_1$ and $\pi_2$ as in the case of non-dependent pairs. But the second projection now uses the self-parameter to guarantee that an element of type $T$ applied to $self.\pi_1$ is returned.

```
data Σ₊(A: Type, T: A -> Type) {
   Pair(A: Type,
        T: A -> Type,
        x: A,
        w: T.ap(A, Type, x) )
      : Σ₊(A, T) }
def Σ₊(A, T).π₁(A: Type, T: A -> Type): A {
   Pair(A, T, x, w) => x }
def (self: Σ₊(A, T)).π₂(A: Type, T: A -> Type)
   : T.ap(A, Type, self.π₁(A, T)) {
   Pair(A, T, x, w) => w }
```

```
codata Σ₋(A: Type, T: A -> Type) {
   Σ₋(A, T).π₁(A: Type, T: A -> Type): A,
   (self: Σ₋(A, T)).π₂(A: Type, T: A -> Type)
      : T.ap(A, Type, self.π₁(A, T)) }
codef Pair(A: Type,
           T: A -> Type,
           x: A,
           w: T.ap(A, Type, x) )
   : Σ₋(A, T) {
   π₁(A, T) => x,
   π₂(A, T) => w }
```

In fact, Agda can already represent Σ-types in both of these ways. But there is one caveat: Agda was not originally designed with codata types in mind, and its codata types are implemented on top of dependent records, which limits what kind of codata types are possible. For example, the order of the destructors in a codata type matter for Agda, so we cannot reorder the first and second projection. In our system the destructors of a codata type are not ordered and can mutually refer to each other, which precisely mirrors how definitions are mutually recursive on the toplevel.

But why should we care about these two alternative encodings of the Σ-type? Take, for example, Eisenberg et al. [2021] who discuss the addition of existential types to Haskell. Since Haskell both is lazy and supports type erasure, Eisenberg et al. are driven to a design that uses strong existential types. We think that by using the framework of data and codata types we can make these kind of differences even clearer.

[3]See the Rust standard library documentation on operators: doc.rust-lang.org/std/ops/index.html.
[4]If we want to define the function type as a data type, then we have to use a system with higher-order inference rules, cf. Garner [2009].

## 2.5 Codata Encodings of Natural Numbers

Starting with the inception of the lambda calculus, researchers have been interested in *functional encodings* of data types such as booleans, natural numbers and lists. Classical examples of functional encodings are the Church, Scott and Parigot encodings of data types (cf. Geuvers [2014]; Koopman et al. [2014]). Functions are from our perspective just one particular instance of a codata type, so we are interested in the more general problem of *codata encodings* instead of functional encodings. Most codata encodings of data types can be obtained by refunctionalizing a data type with an appropriate observation. That the Church encoding can be obtained from refunctionalizing a program with Peano numbers and an `iter` function has already been observed by Ostermann and Jabs [2018]; we restate this example in Figure 6.

```
data Nat { Z, S(p: Nat) }
def Nat.iter(A: Type, z: A, s: A -> A): A {
   Z => z,
   S(p) => s.ap(A, A, p.iter(A, z, s)) }
```

(a) Data variant

```
codata Nat { iter(A: Type, z: A, s: A -> A): A }
codef S(p: Nat): Nat {
   iter(A, z, s) => s.ap(A, A, p.iter(A, z, s)) }
codef Z: Nat { iter(A, z, s) => z }
```

(b) Codata variant

Fig. 6. The Church encoding as a refunctionalized program on Peano numbers.

We can observe that the codata type in Figure 6b which represents the Church encoding of natural numbers is not recursive. This corresponds to the well-known theorem that Church encodings can be typed in pure system F. If we apply the same method to obtain the Scott or Parigot encoding of natural numbers, then we can observe that the resulting codata type is recursive. This corresponds to the other well-known theorem that these encodings can not be typed in pure System F and require recursive types.

We can even go one step further. Geuvers [2001] showed that these previous encodings cannot express induction or dependent elimination. One way to obtain typed functional encodings which can express induction is to add a form of self types to the system; this kind of encoding was introduced by Fu and Stump [2014]. While it is hard to prove an exact correspondence, we think that the essential idea of the encoding of Fu and Stump can be expressed in our system in Figure 7 and Figure 8.

```
codef StepFun(P: Nat -> Type): Fun(Nat, Type) {
   ap(_, _, x) => P.ap(Nat, Type, x) -> P.ap(Nat, Type, S(x)) }
data Nat { S(m: Nat), Z }
def (n: Nat).ind(P: Nat -> Type, base: P.ap(Nat, Type, Z), step: Π(Nat, StepFun(P)))
   : P.ap(Nat, Type, n) {
   S(m) =>
      step.dap(Nat, StepFun(P), m)
          .ap(P.ap(Nat, Type, m), P.ap(Nat, Type, S(m)), m.ind(P, base, step)),
   Z => base }
```

Fig. 7. The data type of natural numbers with an induction principle.

In Figure 7 we have encoded induction using a helper codata type `StepFun` which encodes the induction step for a given predicate $P$ on natural numbers. Induction is then expressed as the observation `ind` on a natural number $n$ which expects the base case and the induction step of the induction as arguments. The argument $n$ on which we define the observation occurs itself in the return type. Refunctionalization of this program results in Figure 8.

```
codef StepFun(P: Nat -> Type): Fun(Nat, Type) {
   ap(_, _, x) => P.ap(Nat, Type, x) -> P.ap(Nat, Type, S(x)) }
codata Nat {
   (n: Nat).ind(P: Nat -> Type, base: P.ap(Nat, Type, Z), step: Π(Nat, StepFun(P)))
      : P.ap(Nat, Type, n) }
codef Z: Nat { ind(P, base, step) => base }
codef S(m: Nat): Nat {
   ind(P, base, step) =>
      step.dap(Nat, StepFun(P), m)
          .ap(P.ap(Nat, Type, m), P.ap(Nat, Type, S(m)), m.ind(P, base, step)) }
```

Fig. 8. The encoding of Fu and Stump can be obtained by refunctionalizing the program in Figure 7.

We think that this is further evidence that the self-parameters we introduced to the system occur naturally when we go from the non-dependent to the dependent setting.

## 3 CASE STUDY

We will now further illustrate the benefits of dependently typed object-oriented programming in a small case study. For this, we create a mockup of a dependently typed web server. We will observe that we can conveniently extend both the supported routes of the web server and the supported methods to access these routes. We will also see how we can conveniently state and enforce properties in intrinsic as well as in extrinsic style.

### 3.1 A Functional Web Server

We start in the familiar realm of functional programming. For the purpose of this demonstration, we will create a simple web server that allows all users to read, but only authenticated users to increment a counter. For this, we track user sessions using the `State` type shown below. As an instance of intrinsic verification, we track on the type level whether the user is authenticated. Possible responses from the server are specified by the `Response` type.

```
codata User { hasCredentials: Bool }
codata State(loggedIn: Bool) {
   State(False).login(u: User): State(u.hasCredentials),
   State(True).logout: State(False),
   State(True).increment: State(True),
   State(True).set(n: Nat): State(True),
   State(b).counter(b: Bool): Nat }
data Response { Forbidden, Return(n: Nat) }
```

Our web server should accept a couple of HTTP request methods (`get`, `post`, ...) for a set of routes (`Index`, `Admin`, ...).

```
data Route { Index }
def Route.requiresLogin: Bool { Index => False }
def (self: Route).get: State(self.requiresLogin) -> Response {
   Index => \state. Return(state.counter(False)) }
```

Adding support for a new request method is as simple as adding a function. For instance, we want to handle post requests, even though we forbid them for the `Index` route:

```
def (self: Route).post: State(self.requiresLogin) -> ×₋(State(self.requiresLogin), Response) {
   Index =>
      \state. comatch {
         fst(a, b) => state,
         snd(a, b) => Forbidden } }
```

### 3.2 Adding New Routes in Object-Oriented Style

While adding new methods is a local change, adding a new route in the functional representation requires touching all pattern matches on `Route` in the program. Therefore, before adding a route to increment the counter on a `POST` request, let us refunctionalize `Route` to its object-oriented decomposition:

```
codata Route {
   requiresLogin: Bool,
   (self: Route).get: State(self.requiresLogin) -> Response,
   (self: Route).post: State(self.requiresLogin) -> ×₋(State(self.requiresLogin), Response) }
codef Index: Route {
   requiresLogin => False,
   get => \state. Return(state.counter(False)),
   post =>
      \state. comatch {
         fst(a, b) => state,
         snd(a, b) => Forbidden } }
```

In the object-oriented decomposition, adding the following `Admin` route is now a local change. Note that the rearrangement works both ways: we could transpose the program back into functional decomposition to add another method.

```
codef Admin: Route {
   requiresLogin => True,
   post =>
      \state. comatch {
         fst(a, b) => state.increment,
         snd(a, b) => Return(state.increment.counter(True)) },
   get => \state. Return(state.counter(True)) }
```

Similar problems of modularity appear in many applications. Functional languages force us to always choose the same extensibility dimension for every type: We can extend data types with new observations, but we cannot easily extend types with new constructors. If the programming language would support both programming paradigms equally well, this choice would not be forced on the programmer by the language, but the programmer would have the choice for each type.

### 3.3 Verifying Properties on Routes

In addition to the ability to increment the counter by sending a `POST` request to the `Admin` route, we may also want to allow explicitly setting a counter value. Updating the counter to a value should be idempotent, i.e. calling the route more than once should have the same effect. The HTTP `PUT` method is supposed to capture this behavior, but how can we enforce it in our code? This leads us to another benefit of the object-oriented style in that we can express such properties extrinsically but still as part of the interface (compare 2.2):

```
data Utils { MkUtils }
def Utils.put_twice(route: Route, request: Request, state: State): Pair(State, Response) {
   MkUtils => route.put(request, route.put(request, state).fst(State, Response)) }
codata Route {
   (self: Route).put(request: Request, state: State): Pair(State, Response),
   (self: Route).put_idempotent(request: Request, state: State)
      : Eq(Pair(State, Response), self.put(request, state), MkUtils.put_twice(self, request, state)) }
```

The full code of the case study with the added `put` method is listed in Appendix B.

### 3.4 The Proof Expression Problem

The classical example of the expression problem [Wadler 1998] concerns extending implementations of term languages by new constructors as well as functions on expressions like *print* or *eval*. Very similar problems arise when formalizing programming languages in proof assistants, where one might want to extend the formalization both by new syntax and by new theorems. In Appendix C, we explore how the proof expression problem manifests differently in an object-oriented proof assistant as opposed to a functional one.

## 4 DESIGN CONSTRAINTS AND SOLUTIONS

Specifying a consistent set of typing and computation rules for data and codata types is not difficult. In this section, we show the difficulties that arise if we also want the rules to be closed under defunctionalization and refunctionalization. That is every program that typechecks should continue to typecheck if we defunctionalize or refunctionalize any of the types that occur in it.

### 4.1 Judgmental Equality of Comatches

*Problem.* For most dependently typed languages, the term $\lambda x.x$ is judgmentally equal to the term $\lambda y.y$, and likewise $\lambda x.2 + 2$ and $\lambda x.4$ are considered equal. Equating such terms becomes a problem, however, if we want to defunctionalize the programs which contain them. Different lambda abstractions in a program are defunctionalized to different constructors, which are then no longer judgmentally equal. Let us illustrate the problem with an example.

Consider the following proof that $\lambda y.y$ is the same function from natural numbers to natural numbers as $\lambda z.z$. We prove this fact using a third lambda abstraction $\lambda x.x$ as an argument to the reflexivity constructor.

```
codata Fun(a b: Type) { Fun(a, b).ap(a b: Type, x: a): b }
Refl(Fun(Nat, Nat), \x. x) : Eq(Fun(Nat, Nat), \y. y, \z. z)
```

If we defunctionalize this program, then each of these three lambda abstractions becomes one constructor of the data type. However since different constructors are not judgmentally equal, the following defunctionalized program no longer typechecks.

```
data Fun(a b: Type) { F1: Fun(Nat, Nat), F2: Fun(Nat, Nat), F3: Fun(Nat, Nat) }
def Fun(a, b).ap(a b: Type, x: a): b { F1 => x, F2 => x, F3 => x }
Refl(Fun(Nat, Nat), F1) : Eq(Fun(Nat, Nat), F2, F3)
```

Here is the gist of the problem: *Judgmental equality must be preserved by defunctionalization and refunctionalization.* This means that if we don't want to treat different constructors of a data type as judgmentally equal, then *we cannot treat all $\alpha$-$\beta$-equivalent comatches as judgmentally equal* either.

It is not impossible to devise a scheme which lifts judgmentally equal comatches to the same constructors. However, we decided against this as it leads to confusing behavior. First, de- and refunctionalization would no longer be inverse transformations at least under syntactic equality. Second, such an attempt would necessarily be a conservative approximation as program equivalence is undecidable in general. In practice, that would mean that some comatches would be collapsed to the same constructor during lifting, while others would not.

*Solution.* Note that the opposite approach—never equating any comatches—doesn't work either, since typing would then no longer be closed under substitution. For example, if $f$ is a variable standing for a function from natural numbers to natural numbers, then the term $\mathrm{Refl}(\mathrm{Fun}(\mathrm{Nat}, \mathrm{Nat}), f)$ is a proof of the proposition $\mathrm{Eq}(\mathrm{Fun}(\mathrm{Nat}, \mathrm{Nat}), f, f)$. But we could not substitute a comatch $\lambda y.y$ for $f$, since the result would no longer typecheck. We therefore have to find a solution between these two extremes.

Our solution consists of always considering local comatches together with a name[5]. Only comatches which have the same name are judgmentally equal, and this equality is preserved by reduction since the comatch is duplicated together with its name.

Where do the names for local comatches come from? We support user-annotated labels, which allow the programmer to give meaningful names to comatches. Manually naming comatches in this way is useful as these labels can also be used by defunctionalization to name the generated constructors. We enforce that these user-annotated labels are globally unique. However, as we do not want to burden the user with naming every single comatch in the program, we also allow unannotated comatches, for which we automatically generate unique names. As a result, each comatch occurring textually in the program has a unique name, but these names possibly become duplicated during normalization and typechecking.

[5]This solution is similar to Binder et al.'s use of labels for local (co)pattern matches

### 4.2 Eta Equality

*Problem.* For reasons very similar to the previous section, $\eta$-equality is not preserved under defunctionalization and refunctionalization. Let us again consider a simple example. In the following proof, we show that a function $f$ is equal to its $\eta$-expanded form $\lambda x.f.\texttt{ap}(x)$. In order to typecheck, the proof would need to use a judgmental $\eta$-equality for functions.

```
codata Fun { ap(x: Nat): Nat }
let prop_eta(f: Fun): Eq(Fun, f, (\x. f.ap(x))) := Refl(Fun, f);
```

However, defunctionalization of this proof would result in the following program, where we have used an ellipsis to mark all the constructors that were generated for the other lambda abstractions in the program.

```
data Fun { Eta(f: Fun), ... }
def Fun.ap(x: Nat): Nat { Eta(f) => f.ap(x),... }
let prop_eta(f: Fun): Eq(Fun, f, Eta(f)) := Refl(Fun, f);
```

Using `prop_eta` it would now be possible to show that any constructor `f` of `Fun` is equal to `Eta(f)`. This would contradict the provable proposition that distinct constructors are not equal.

*Solution.* We do not support $\eta$-equality in our formalization and implementation. This means that we only normalize $\beta$-redexes but not $\eta$-redexes during typechecking. However, it would be possible to support judgmental $\eta$-equality on a case-by-case basis similar to the `eta-equality` and `no-eta-equality` keywords in Agda which enable or disable `eta`-equality for a specific record type[6]. De- and refunctionalization is then only available for types without $\eta$-equality.

## 5 FORMALIZATION

In this section, we present the syntax, typing rules and operational semantics of our system. We divide this presentation into three subsections: In Section 5.1, we introduce the core of our system. We extend this core calculus by data types and pattern matching definitions in Section 5.2, and by codata types and copattern matching definitions in Section 5.3.

We do not formalize local pattern and copattern matches. Instead, local pattern and copattern matches are lifted to the top level before applying de- or refunctionalization, similar to the approach taken by Binder et al. [2019]. Some care must be taken to ensure that we close over all required terms, as the types of terms which are part of the closure might close over additional terms. For example, closing over `v: Vec n` requires us to also close over `n`. The main challenge for local pattern and copattern matches revolves around judgmental equality, which we discussed in Section 4.1.

### 5.1 Core System

In Figure 9 we define the syntax of our core system together with small examples in the rightmost column.

[6]Compare the section on record types in the Agda user manual: agda.readthedocs.io/en/v2.6.3/language/record-types.html.

| | | | | |
|---|---|---|---|---|
| $\mathcal{T}$ | $\in$ | TypeNames | *Type names* | `Bool`, `Vec`, `Stream` |
| C | $\in$ | ProducerNames | *Producer names* | `True`, `Cons` |
| d | $\in$ | ConsumerNames | *Consumer names* | `neg`, `neg_inverse` |
| x, y, z | $\in$ | Variables | *Variables* | |
| $\delta$ | ::= | (empty in core system) | *Declaration* | |
| $\Theta$ | ::= | $\emptyset \mid \delta, \Theta$ | *Program* | |
| $\Gamma, \Delta$ | ::= | $\emptyset \mid \Gamma, \mathrm{x} : t$ | *Context* | x : `Nat`, v : `Vec(Bool, x)` |
| $\Xi, \Psi$ | ::= | $\emptyset \mid \mathrm{x} : t, \Xi$ | *Telescope* | x : `Nat`, v : `Vec(Bool, x)` |
| $\rho, \sigma$ | ::= | $() \mid (e, \sigma)$ | *Substitution* | `(Bool, S(Z))` |
| $e, s, t$ | ::= | x | *Variable* | |
| | $\mid$ | Type | *Universe* | |
| | $\mid$ | $\mathcal{T}\rho$ | *Type* | `Bool`, `Vec(Bool, S(Z))` |
| | $\mid$ | $\mathrm{C}\sigma$ | *Producer* | `S(Z)` |
| | $\mid$ | $e.\mathrm{d}\sigma$ | *Consumer* | `x.neg` |

Fig. 9. Syntax of core system without data or codata types.

Following standard convention, we formalize our system up to $\alpha$-renaming of bound variables x, y, z. We distinguish between contexts $\Gamma$, $\Delta$ and telescopes $\Xi$, $\Psi$. Contexts track the types of free variables and must always be closed. Telescopes are dependent parameter lists whose types may contain free variables bound in a context. If a telescope is closed, we may implicitly use it as a context. A substitution $\rho$, $\sigma$ is an argument list to a telescope. A program $\Theta$ is a list of declarations $\delta$, which are empty for now. There are five different kinds of expressions $e$, $s$, $t$: Variables are denoted as described above. We denote the type universe as Type. Type constructors $\mathcal{T}\rho$ instantiate a (co)data type with a substitution $\rho$. Calling a producer C is written $\mathrm{C}\sigma$; invoking a consumer d uses the syntax $e.\mathrm{d}\sigma$. The producer syntax denotes constructor calls for data types and codefinition calls for codata types. The consumer syntax denotes destructor invocations for codata and definition calls for data types.

For formalizing the core system, we closely follow the presentation by Hofmann [1997]. The rules for contexts, telescopes, and substitutions are standard, and we omit them here for space reasons. We will use the judgment forms $\vdash_\Theta \Gamma$ ctx and $\Gamma \vdash_\Theta \Xi$ tel to specify valid contexts and telescopes, respectively. The judgment form $\Gamma \vdash_\Theta \sigma : \Xi$ states that $\sigma$ is a valid substitution for the telescope $\Xi$ under context $\Gamma$.

We also assume the Type-in-Type axiom, which is well-known to be inconsistent [Girard 1972; Hurkens 1995]. In this work, we do not enforce termination or productivity in our system. As adding the Type-in-Type axiom merely yields another source of possible divergence [Tennant 1982], we do not gain anything from avoiding this paradox, e.g. by using a hierarchy of universes. We therefore follow Eisenberg [2016] and opt for a simpler presentation using the Type-in-Type axiom.

## 5.2 Data Types and Dependent Pattern Matching

We now extend the declarations of Figure 9 by two new constructs: data type declarations and pattern matching definitions. Data type declarations introduce a new data type together with a list of constructors, and pattern matching definitions introduce a top-level consumer which is defined by a list of clauses. The corresponding new typing and well-formedness rules are contained in the upper half of Figure 10.

$$\begin{array}{rcllll} \delta & ::= & \ldots & \textit{Extends Figure 9} & \\ & | & \textbf{data}\ \mathcal{T}\Psi\ \{\ \overline{\mathrm{C}\Xi : \mathcal{T}\rho}\ \} & \textit{Data Type} & \textbf{data}\ \texttt{Bool}\ \{\ \texttt{True} : \texttt{Bool}, \ldots \} \\ & | & \textbf{def}\ (\mathrm{z} : \mathcal{T}\rho).\mathrm{d}\Xi : t\ \{\ \overline{a}\ \} & \textit{Pattern Match} & \textbf{def}\ (\mathrm{x} : \texttt{Bool}).\texttt{neg} : \texttt{Bool}\ \{\ldots\} \\ a & ::= & \mathrm{C}\ \Xi \mapsto e & \textit{Match case} & \texttt{S}(\mathrm{x} : \texttt{Nat}) \mapsto \mathrm{x} \\ & | & \mathrm{C}\ \Xi\ \textbf{absurd} & \textit{Absurd case} & \texttt{Nil}\ (\mathrm{a} : \mathrm{Type})\ \textbf{absurd} \end{array}$$

A data type declaration **data** $\mathcal{T}\Psi\ \{\ \overline{\mathrm{C}\Xi : \mathcal{T}\rho}\ \}$ introduces one type constructor $\mathcal{T}$ and a list of term constructors C to the rest of the program. The type constructor $\mathcal{T}$ is indexed by a telescope $\Psi$, and if we provide a substitution $\rho$ for this telescope, then the formation rule F-Data allows us to form the type $\mathcal{T}\rho$. Each term constructor is declared with parameters $\Xi$ and a type $\mathcal{T}\rho$ which specifies how the term constructor instantiates the indices of the type constructor. We can use a term constructor with the introduction rule I-Data, where the resulting type depends on both the arguments $\sigma$ to which the term constructor is applied and the substitution $\rho$ from the constructor declaration. We check that a data type declaration is well-formed with the help of the rule Data: For the type constructor $\mathcal{T}$ we check that the indices $\Psi$ form a valid telescope in the program, and for each constructor $\mathrm{C}\Xi : \mathcal{T}\rho$ we check that $\Xi$ is a valid telescope, and that $\rho$ is a valid substitution for the indices $\Psi$ of the type constructor $\mathcal{T}$ under context $\Xi$.

The second new construct is pattern matching definitions **def** $(\mathrm{z} : \mathcal{T}\rho).\mathrm{d}\Xi : t\ \{\ \overline{a}\ \}$ which define a new consumer d by a list of cases. The consumer d takes arguments specified by the telescope $\Xi$, and can be called on any term of type $\mathcal{T}\rho$, where the substitution $\rho$ can depend on any arguments bound in $\Xi$. The scrutinee of the consumer can be referred to by the self parameter z in the return type $t$ of d. We introduced these self-parameters in Section 2.1. The elimination rule E-Data then eliminates a term $e$ by invoking d with arguments $\sigma$. These arguments are substituted in the return type $t$ which is defined under the context $\Xi; z : \mathcal{T}\rho$. Here, the self parameter $z$ is replaced by the scrutinee $e$.

We check whether a pattern matching definition is well-formed with the rule Def. The return type $t$ is typed under context $\Xi; \mathrm{z} : \mathcal{T}\rho$. We implicitly require that there exists exactly one case of each constructor C of $\mathcal{T}$, and check that every case is well-formed. For checking the well-formedness of cases we use an auxiliary judgment form $\Xi \vdash_\Theta a_i : (z : \mathcal{T}\rho).t$ which tracks the self parameter $\mathrm{z} : \mathcal{T}\rho$ and the return type $t$. There are two variants of pattern matching cases we have to consider: possible and absurd cases. In a possible case, we restate the constructor telescope $\Xi$ and give an expression $e$ that gives the result if the case is matched. An absurd case is determined to be impossible by unification and hence does not need an expression. Corresponding to the two kinds of cases, possible and absurd, there are two typing rules, $\textsc{Case}_1$ and $\textsc{Case}_2$. In both rules, we unify the scrutinee type $\mathcal{T}\rho_1$ with the constructor type $\mathcal{T}\rho_2$. If there is no such unifier, the case is absurd. In a possible case, however, unification yields a unifier $\theta$. We use this unifier to refine the typing of the right-hand side $e$: The unifier $\theta$ is substituted in the context $\Xi_1; \Xi_2$, expression $e$ and type $t$. We further refine $t$ by replacing the self parameter z with the constructor $\mathrm{Cid}_{\Xi_2}$ where $\mathrm{id}_{\Xi_2}$ is the identity context morphism for $\Xi_2$.

Finally, the computation rule ≡-Data allows us to reduce an expression $\mathrm{C}\sigma_1.\mathrm{d}\sigma_2$ if $\sigma_1$ and $\sigma_2$ are valid arguments to the constructor respectively definition.

### 5.3 Codata Types and Dependent Copattern Matching

Finally, we extend programs by codata type declarations and copattern matching definitions. Codata type declarations introduce a new codata type together with a list of destructors, and copattern matching definitions introduce a new top-level producer by a list of copattern matching

**Declaration Rules**

$$\frac{\vdash_\Theta \Psi\ \text{tel} \qquad \forall i : \left[\ \vdash_\Theta \Xi_i\ \text{tel}\ \text{ and }\ \Xi_i \vdash_\Theta \rho_i : \Psi \right]}{\vdash_\Theta \textbf{data}\ \mathcal{T}\Psi\ \{\ \overline{\mathrm{C}\Xi : \mathcal{T}\rho}\ \}\ \textsc{Ok}}\ \textsc{Data}$$

$$\frac{\textbf{data}\ \mathcal{T}\Psi\ \{...\} \in \Theta \quad \Xi; \mathrm{z} : \mathcal{T}\rho \vdash_\Theta t : \mathrm{Type} \qquad \Xi \vdash_\Theta \rho : \Psi \quad \forall i :\ \Xi \vdash_\Theta a_i : (\mathrm{z} : \mathcal{T}\rho).t}{\vdash_\Theta \textbf{def}\ (\mathrm{z} : \mathcal{T}\rho).\mathrm{d}\Xi : t\ \{\ \overline{a}\ \}\ \textsc{Ok}}\ \textsc{Def}$$

$$\frac{\textbf{data}\ \mathcal{T}\Psi\ \{\ \mathrm{C}\Xi_2 : \mathcal{T}\rho_2,...\} \in \Theta \qquad \Xi_1;\Xi_2 \vdash_\Theta \theta\ \text{mgu for}\ \rho_1 \equiv \rho_2 : \Psi \qquad (\Xi_1;\Xi_2)[\theta] \vdash_\Theta e[\theta] : t[\mathrm{C}\ \mathrm{id}_{\Xi_2}/\mathrm{z}][\theta]}{\Xi_1 \vdash_\Theta \mathrm{C}\ \Xi_2 \mapsto e : (\mathrm{z} : \mathcal{T}\rho_1).t}\ \textsc{Case}_1$$

$$\frac{\textbf{data}\ \mathcal{T}\Psi\ \{\ \mathrm{C}\Xi_2 : \mathcal{T}\rho_2,...\} \in \Theta \qquad \neg\exists\theta :\ \Xi_1;\Xi_2 \vdash_\Theta \theta\ \text{mgu for}\ \rho_1 \equiv \rho_2 : \Psi}{\Xi_1 \vdash_\Theta \mathrm{C}\ \Xi_2\ \textbf{absurd} : (\mathrm{z} : \mathcal{T}\rho_1).t}\ \textsc{Case}_2$$

**Formation, Introduction, Elimination and Computation Rules**

$$\frac{\textbf{data}\ \mathcal{T}\Psi\ \{...\} \in \Theta \qquad \Gamma \vdash_\Theta \rho : \Psi}{\Gamma \vdash_\Theta \mathcal{T}\rho : \mathrm{Type}}\ \textsc{F-Data}$$

$$\frac{\textbf{data}\ \mathcal{T}\Psi\ \{\ \mathrm{C}\Xi : \mathcal{T}\rho,...\} \in \Theta \qquad \Gamma \vdash_\Theta \sigma : \Xi}{\Gamma \vdash_\Theta \mathrm{C}\sigma : \mathcal{T}\rho[\sigma/\Xi]}\ \textsc{I-Data}$$

$$\frac{\textbf{def}\ (z : \mathcal{T}\rho).\mathrm{d}\Xi : t\ \{...\} \in \Theta \qquad \Gamma \vdash_\Theta \sigma : \Xi \qquad \Gamma \vdash_\Theta e : \mathcal{T}\rho[\sigma/\Xi]}{\Gamma \vdash_\Theta e.\mathrm{d}\sigma : t[\sigma/\Xi][e/\mathrm{z}]}\ \textsc{E-Data}$$

$$\frac{\textbf{data}\ \mathcal{T}\Psi\ \{\ \mathrm{C}\Xi_1 : \mathcal{T}\rho_1,...\} \in \Theta \quad \Gamma \vdash_\Theta \sigma_1 : \Xi_1 \qquad \textbf{def}\ (\mathrm{z} : \mathcal{T}\ \rho_2).\mathrm{d}\ \Xi_2 : t\ \{\ \mathrm{C}\ \Xi_1 \mapsto e,...\ \} \in \Theta \quad \Gamma \vdash_\Theta \sigma_2 : \Xi_2 \qquad \Gamma \vdash_\Theta \rho_1[\sigma_1/\Xi_1] \equiv \rho_2[\sigma_2/\Xi_2] : \Psi}{\Gamma \vdash_\Theta \mathrm{C}\sigma_1.\mathrm{d}\sigma_2 \equiv e[\sigma_2/\Xi_2][\sigma_1/\Xi_1] : t[\sigma_2/\Xi_2][\mathrm{C}\sigma_1/\mathrm{z}]}\ \equiv\textsc{-Data}$$

---

**Declaration Rules**

$$\frac{\vdash_\Theta \Psi\ \text{tel} \qquad \forall i : \begin{bmatrix} \vdash_\Theta \Xi_i\ \text{tel}\ \text{ and} \\ \Xi_i \vdash_\Theta \rho_i : \Psi\ \text{ and} \\ \Xi_i; \mathrm{z}_i : \mathcal{T}\rho_i \vdash_\Theta t_i : \mathrm{Type} \end{bmatrix}}{\vdash_\Theta \textbf{codata}\ \mathcal{T}\ \Psi\ \{\ \overline{(\mathrm{z} : \mathcal{T}\rho).\ \mathrm{d}\ \Xi : t}\ \}\ \textsc{Ok}}\ \textsc{Codata}$$

$$\frac{\textbf{codata}\ \mathcal{T}\Psi\{...\} \in \Theta \qquad \Xi \vdash_\Theta \rho : \Psi \qquad \forall i, \Xi \vdash_\Theta o_i : (C : \mathcal{T}\rho)}{\vdash_\Theta \textbf{codef}\ \mathrm{C}\ \Xi : \mathcal{T}\ \rho\ \{\ \overline{o}\ \}\ \textsc{Ok}}\ \textsc{Codef}$$

$$\frac{\textbf{codata}\ \mathcal{T}\ \Psi\ \{\ (\mathrm{z} : \mathcal{T}\rho_2).\ \mathrm{d}\ \Xi_2 : t\ \} \in \Theta \qquad \Xi_1;\Xi_2 \vdash_\Theta \theta\ \text{mgu for}\ \rho_1 \equiv \rho_2 : \Psi \qquad (\Xi_1;\Xi_2)[\theta] \vdash_\Theta e[\theta] : t[\mathrm{C}\ \mathrm{id}_{\Xi_1}/\mathrm{z}][\theta]}{\Xi_1 \vdash_\Theta \mathrm{d}\ \Xi_2 \mapsto e : (C : \mathcal{T}\ \rho_1)}\ \textsc{Cocase}_1$$

$$\frac{\textbf{codata}\ \mathcal{T}\ \Psi\ \{\ (\mathrm{z} : \mathcal{T}\rho_2).\ \mathrm{d}\ \Xi_2 : t\ \} \in \Theta \qquad \neg\exists\theta, \Xi_1;\Xi_2 \vdash_\Theta \theta\ \text{mgu for}\ \rho_1 \equiv \rho_2 : \Psi}{\Xi_1 \vdash_\Theta \mathrm{d}\ \Xi_2\ \textbf{absurd} : (C : \mathcal{T}\ \rho)}\ \textsc{Cocase}_2$$

**Formation, Introduction, Elimination and Computation Rules**

$$\frac{\textbf{codata}\ \mathcal{T}\Psi\ \{...\} \in \Theta \qquad \Gamma \vdash_\Theta \rho : \Psi}{\Gamma \vdash_\Theta \mathcal{T}\rho : \mathrm{Type}}\ \textsc{F-Codata}$$

$$\frac{\textbf{codef}\ \mathrm{C}\Xi : \mathcal{T}\rho\ \{...\} \in \Theta \qquad \Gamma \vdash_\Theta \sigma : \Xi}{\Gamma \vdash_\Theta \mathrm{C}\sigma : \mathcal{T}\rho[\sigma/\Xi]}\ \textsc{I-Codata}$$

$$\frac{\textbf{codata}\ \mathcal{T}\Psi\ \{(z : \mathcal{T}\rho).\mathrm{d}\Xi : t,...\} \in \Theta \qquad \Gamma \vdash_\Theta \sigma : \Xi \qquad \Gamma \vdash_\Theta e : \mathcal{T}\rho[\sigma/\Xi]}{\Gamma \vdash_\Theta e.\mathrm{d}\sigma : t[\sigma/\Xi][e/\mathrm{z}]}\ \textsc{E-Codata}$$

$$\frac{\textbf{codata}\ \mathcal{T}\Psi\ \{(z : \mathcal{T}\rho_2).\mathrm{d}\Xi_2 : t,...\} \in \Theta \quad \Gamma \vdash_\Theta \sigma_1 : \Xi_1 \qquad \textbf{codef}\ \mathrm{C}\Xi_1 : \mathcal{T}\rho_1\ \{\mathrm{d}\ \Xi_2 \mapsto e,...\} \in \Theta \quad \Gamma \vdash_\Theta \sigma_2 : \Xi_2 \qquad \Gamma \vdash_\Theta \rho_1[\sigma_1/\Xi_1] \equiv \rho_2[\sigma_2/\Xi_2] : \Psi}{\Gamma \vdash_\Theta \mathrm{C}\sigma_1.\mathrm{d}\sigma_2 \equiv e[\sigma_1/\Xi_1][\sigma_2/\Xi_2] : t[\sigma_2/\Xi_2][\mathrm{C}\sigma_1/\mathrm{z}]}\ \equiv\textsc{-Codata}$$

Fig. 10. Well-formedness and typing rules for data and codata types.

clauses. Their typing and well-formedness rules are contained in the lower half of Figure 10.

$$\begin{array}{lcll}
\delta & ::= & \ldots & \textit{Extends Figure 9} \\
 & \mid & \textbf{codata}\ \mathcal{T}\Psi\ \{\ \overline{(\mathrm{z} : \mathcal{T}\rho).\mathrm{d}\Xi : t}\ \} & \textit{Codata declaration} \\
 & \mid & \textbf{codef}\ \mathrm{C}\Xi : \mathcal{T}\rho\ \{\ \overline{o}\ \} & \textit{Producer declaration} \\
 & & & \\
o & ::= & \mathrm{d}\ \Xi \mapsto e & \textit{Possible cocase} \\
 & \mid & \mathrm{d}\ \Xi\ \textbf{absurd} & \textit{Absurd cocase}
\end{array}$$

A codata type declaration $\textbf{codata}\ \mathcal{T}\Psi\ \{\ \overline{(\mathrm{z} : \mathcal{T}\rho).\mathrm{d}\Xi : t}\ \}$ introduces the type constructor $\mathcal{T}$ and a list of destructors d to the program. Like data types, codata types are indexed, and the type constructor $\mathcal{T}$ can be instantiated to form types with the help of the rule F-Codata. The signature of each destructor declaration $(z : \mathcal{T}\rho).\mathrm{d}\Xi : t$ corresponds precisely to the signature of a pattern matching definition for data types: Destructors take a parameter telescope $\Xi$ and a self parameter $(z : \mathcal{T}\rho)$, where $\rho$ is a substitution for the type parameters $\Psi$ under context $\Xi$. The return type $t$ is defined under the context $\Xi; z : \mathcal{T}\rho$. Using the rule E-Codata, we can eliminate a term $e$ by calling a destructor d with arguments $\sigma$. In order to obtain the type of $e.\mathrm{d}\sigma$, we substitute $\sigma$ for the parameters $\Xi$ in the return type $t$, and $e$ for the self parameter z. We use the rule Codata to check that a codata type declaration is well-formed.

A codefinition $\textbf{codef}\ \mathrm{C}\Xi : \mathcal{T}\rho\ \{\ \overline{o}\ \}$ has a signature which reflects the signature of constructors of a data type. The body of a codefinition is a list of *cocases*, which can be either possible or absurd. In a possible cocase, we restate the telescope $\Xi$ and give an expression $e$ which provides the result if the cocase gets matched. An absurd cocase is determined to be impossible by unification and hence does not need to provide an expression.

We check whether codefinitions are well-formed with the help of the rule Codef: We check that the parameters $\Xi$ are valid and that the arguments $\rho$ to the type constructor are well-typed under context $\Xi$. Further, we ensure that all cocases $o_i$ are well-formed with the auxiliary judgment form $\Xi \vdash_\Theta m_i^d : (C : \mathcal{T}\ \rho)$, which tracks label C and type $\mathcal{T}\ \rho$ of the codefinition. We implicitly require that there exists one cocase for each destructor d of $\mathcal{T}$. As with pattern matching cases for data types, there are possible and absurd cocases as specified by the rules $\textsc{Cocase}_1$ and $\textsc{Cocase}_2$. Which rule applies is determined by unifying the codefinition type $\mathcal{T}\rho_1$ with the destructor type $\mathcal{T}\rho_2$. If no such unifier exists, the cocase is absurd. In a possible cocase, the unifier $\theta$ is substituted in context $\Xi_1; \Xi_2$, expression $e$, and type $t$. We also refine the return type $t$ by substituting $\mathrm{C}\ \mathrm{id}_{\Xi_1}$, where $\mathrm{id}_{\Xi_1}$ is the identity context morphism for $\Xi_1$. Reminiscent of constructor calls, I-Codata introduces a term of such a type by invoking a codefinition C with arguments $\sigma$ for the parameters $\Xi$. As these parameters $\Xi$ may occur in the constructed type $\mathcal{T}\ \rho$, we substitute the arguments $\sigma$ in the type.

Lastly, the computation rule $\equiv$-Codata reduces a redex $\mathrm{C}\sigma_1.\mathrm{d}\sigma_2$ if $\sigma_1$ is a valid argument for the codefinition C and $\sigma_2$ is a valid argument for the destructor d.

## 5.4 Call-By-Value Operational Semantics

The computation rules of Section 5.2 and Section 5.3 do not specify a deterministic evaluation order. In this section we present the call-by-value (CBV) operational semantics of our system; the operational semantics does not depend on typing information. We specify evaluation by means of evaluation contexts in the style of Felleisen and Hieb [1992]. Values consist of the type universe or of type constructors and named producers applied to other values. Named producers can either be the constructors of a data type or the call to a toplevel codefinition. Such codefinitions generalize lambda abstractions which are lifted to the toplevel.

$$\begin{array}{lcl}
v & ::= & \mathrm{Type} \mid \mathcal{T}\,\overline{v} \mid \mathrm{C}\,\overline{v} \\
E & ::= & \square \mid \mathrm{C}\,\overline{v} E\,\overline{e} \mid \mathcal{T}\,\overline{v} E\,\overline{e} \mid E.\mathrm{d}\sigma \mid v.d\,\overline{v} E\,\overline{e}
\end{array}$$

Reduction $e \triangleright e'$ happens when introduction and elimination forms meet, i.e. when a method is called on a constructor, or when a destructor is invoked on a codefinition:

$$\frac{e \triangleright_\beta e'}{E[e] \triangleright E[e']} \qquad \frac{\textbf{def } (z : \mathcal{T}\rho).\mathrm{d}\Xi_2 : t\ \{\mathrm{C}\ \Xi_1 \mapsto e\} \in \Theta}{\mathrm{C}\overline{v_1}.\mathrm{d}\overline{v_2} \triangleright_\beta e[\overline{v_2}/\Xi_2][\overline{v_1}/\Xi_1]} \qquad \frac{\textbf{codef } \mathrm{C}\Xi_1 : \mathcal{T}\rho_1\ \{\mathrm{d}\ \Xi_2 \mapsto e\} \in \Theta}{\mathrm{C}\overline{v_1}.\mathrm{d}\overline{v_2} \triangleright_\beta e[\overline{v_1}/\Xi_1][\overline{v_2}/\Xi_2]}$$

Here, $\triangleright_\beta$ denotes single step reduction of a direct redex, while $\triangleright$ denotes evaluation within an evaluation context.

### 5.5 Type Soundness

We show type soundness with respect to the call-by-value semantics of Section 5.4 by the usual progress and preservation theorems.

Theorem 5.1 (Progress). *For any well-formed program $\Theta$ and expressions $e$ and $t$, if $\vdash_\Theta e : t$ then either $e$ is a value or there exists an expression $e'$ such that $e \triangleright e'$.*

Proof. See Appendix A.1. □

Theorem 5.2 (Preservation). *For any well-formed program $\Theta$ and any expressions $e_1, e_2, t$ if $\vdash_\Theta e_1 : t$ and $e_1 \triangleright e_2$, then $\vdash_\Theta e_2 : t$.*

Proof. See Appendix A.1. □

## 6 DE/REFUNCTIONALIZATION

In our system, de- and refunctionalization can transform any data type into a codata type and vice versa. The key insight behind these transformations is that any data or codata type in the program can be represented in matrix form as shown in Figure 11. A data type is fully determined by specifying an expression for each constructor-definition pair, while a codata type is fully determined by giving an expression for each destructor-codefinition pair. Using these matrix representations, the process of de- and refunctionalization simplifies to matrix transposition. With this intuition in mind, we will now formally define de- and refunctionalization and state our main propositions.

| Ctor \ Def | $(z_1 : \mathcal{T}\rho_1).\mathrm{d}_1\Xi_1 : t_1$ | $\cdots$ | $(z_m : \mathcal{T}\rho_m).\mathrm{d}_m\Xi_m : t_m$ |
|---|---|---|---|
| $\mathrm{C}_1\Xi_1 : \mathcal{T}\rho_1$ | $e_{1,1}$ | $\cdots$ | $e_{1,m}$ |
| $\vdots$ | $\vdots$ | | $\vdots$ |
| $\mathrm{C}_n\Xi_n : \mathcal{T}\rho_n$ | $e_{n,1}$ | $\cdots$ | $e_{n,m}$ |

(a) The data matrix.

| Dtor \ Codef | $\mathrm{C}_1\Xi_1 : \mathcal{T}\rho_1$ | $\cdots$ | $\mathrm{C}_n\Xi_n : \mathcal{T}\rho_n$ |
|---|---|---|---|
| $(z_1 : \mathcal{T}\rho_1).\mathrm{d}_1\Xi_1 : t_1$ | $e_{1,1}$ | $\cdots$ | $e_{n,1}$ |
| $\vdots$ | $\vdots$ | | $\vdots$ |
| $(z_m : \mathcal{T}\rho_m).\mathrm{d}_m\Xi_m : t_m$ | $e_{1,m}$ | $\cdots$ | $e_{n,m}$ |

(b) The codata matrix.

Fig. 11. Data and codata matrix.

*Definition 6.1 (Defunctionalization).* We write $\mathcal{D}_\mathcal{T}(\Theta)$ to denote the defunctionalization of $\mathcal{T}$ in $\Theta$. The precondition for applying $\mathcal{D}_\mathcal{T}(\Theta)$ is that **codata** $\mathcal{T}\ \Psi\ \{\ ...\ \} \in \Theta$. Defunctionalization transforms the codata type into a data type. For each destructor d and each codefinition C in $\Theta$:

$$\textbf{codata } \mathcal{T}\ \Psi\ \{\ (\mathrm{z} : \mathcal{T}\rho_2).\ \mathrm{d}\ \Xi_2 : t, ...\ \}, \textbf{codef } \mathrm{C}\ \Xi_1 : \mathcal{T}\ \rho_1\ \{\ \mathrm{d}\ \Xi_2 \mapsto e, ...\ \}$$

the program $\mathcal{D}_\mathcal{T}(\Theta)$ contains a corresponding constructor C and a definition d:

$$\textbf{data } \mathcal{T}\ \Psi\ \{\ \mathrm{C}\ \Xi_1 : \mathcal{T}\ \rho_1, ...\ \}, \textbf{def } (\mathrm{z} : \mathcal{T}\ \rho_2).\mathrm{d}\ \Xi_2 : t\ \{\ \mathrm{C}\ \Xi_1 \mapsto e, ...\ \}$$

*Definition 6.2 (Refunctionalization).* We write $\mathcal{R}_\mathcal{T}(\Theta)$ to denote the refunctionalization of $\mathcal{T}$ in $\Theta$. The precondition for applying $\mathcal{R}_\mathcal{T}(\Theta)$ is that **data** $\mathcal{T}\ \Psi\ \{\ ...\ \} \in \Theta$. Refunctionalization transforms the data type into a codata type. For each constructor C and each definition d in $\Theta$:

$$\textbf{data } \mathcal{T}\ \Psi\ \{\ \mathrm{C}\ \Xi_1 : \mathcal{T}\ \rho_1, ...\ \}, \textbf{def } (\mathrm{z} : \mathcal{T}\ \rho_2).\mathrm{d}\ \Xi_2 : t\ \{\ \mathrm{C}\ \Xi_1 \mapsto e, ...\ \}$$

the program $\mathcal{R}_{\mathcal{T}}(\Theta)$ contains a corresponding codefinition C and a destructor d:

$$\textbf{codata}\ \mathcal{T}\ \Psi\ \{\ (\mathrm{z} : \mathcal{T}\rho_2).\ \mathrm{d}\ \Xi_2 : t, \ldots\ \}, \textbf{codef}\ \mathrm{C}\ \Xi_1 : \mathcal{T}\ \rho_1\ \{\ \mathrm{d}\ \Xi_2 \mapsto e, \ldots\ \}$$

We write $\mathcal{X}_{\mathcal{T}}(\cdot)$ for both $\mathcal{D}_{\mathcal{T}}(\cdot)$ and $\mathcal{R}_{\mathcal{T}}(\cdot)$, when their difference doesn't matter. Using these definitions, we can now state our main propositions. Notice that de- and refunctionalization do not affect the program on the expression level. This is because we reuse the syntax for *producers* for both constructor and codefinition calls and the syntax for *consumers* for both destructor and definition calls (see Figure 9). Therefore, in the proposition statements below, de-/refunctionalization is only applied to the program $\Theta$.

Theorem 6.3 (De/Refunctionalization preserves typing and judgmental equality). *The following implications hold:*

- $\Gamma \vdash_\Theta e : t \implies \Gamma \vdash_{\mathcal{X}_{\mathcal{T}}(\Theta)} e : t$
- $\Gamma \vdash_\Theta e_1 \equiv e_2 : t \implies \Gamma \vdash_{\mathcal{X}_{\mathcal{T}}(\Theta)} e_1 \equiv e_2 : t$
- $\Gamma \vdash_\Theta \sigma : \Xi \implies \Gamma \vdash_{\mathcal{X}_{\mathcal{T}}(\Theta)} \sigma : \Xi$
- $\Gamma \vdash_\Theta \sigma_1 \equiv \sigma_2 : \Xi \implies \Gamma \vdash_{\mathcal{X}_{\mathcal{T}}(\Theta)} \sigma_1 \equiv \sigma_2 : \Xi$
- $\vdash_\Theta \Gamma\ \mathrm{ctx} \implies \vdash_{\mathcal{X}_{\mathcal{T}}(\Theta)} \Gamma\ \mathrm{ctx}$
- $\Gamma \vdash_\Theta \Xi\ \mathrm{tel} \implies \Gamma \vdash_{\mathcal{X}_{\mathcal{T}}(\Theta)} \Xi\ \mathrm{tel}$

Proof. See Appendix A.4 for a proof outline. □

Theorem 6.4 (De/Refunctionalization preserves well-formedness of programs). *If* $\vdash_\Theta \Theta$ Ok, *then* $\vdash_\Theta \mathcal{X}_{\mathcal{T}}(\Theta)$ Ok

Proof. See Appendix A.4 for a proof outline. □

## 7 FUTURE WORK

In this paper, we described a dependently typed programming language based on data and codata. How to extend this programming language to a proof assistant is one of the problems that we want to address in the future. In the following sections, we describe the problems that have to be solved to make our system consistent, in a way that is compatible with the transformations we described.

### 7.1 Specifying Termination and Productivity

The system we presented does not have any form of termination or productivity checking. We could, of course, use any of the existing off-the-shelf solutions for checking termination and productivity. The problem with that approach is that, in general, a program that typechecks and is verified to only have terminating recursive definitions and productive corecursive definitions might not be verifiably total after de/-refunctionalization. We illustrate this with the following example:

```
data Nat { S(x: Nat), Z }
def Nat.plus(n: Nat): Nat {
   Z => n,
   S(x') => S(x'.plus(n)) }
def Nat.mul(n: Nat): Nat {
   Z => Z,
   S(m) => n.plus(m.mul(n)) }
```

```
codata Nat { plus(n: Nat): Nat, mul(n: Nat): Nat }
codef S(x: Nat): Nat {
   plus(n) => S(x.plus(n)),
   mul(n) => n.plus(x.mul(n)) }
codef Z: Nat {
   plus(n) => n,
   mul(n) => Z }
```

We could check termination for the program on the left in the usual way. We check the definition of `plus` first and verify that it is only called on the structurally smaller argument `m`. We then add `plus` to the context of functions which are checked to be total. We then check the definition of `mul`, having the function `plus` as a total function in the context. We see again that `mul` is called on the structurally smaller argument `m`.

Refunctionalizing the program on the left results in the program on the right. In this program, we have to check the productivity of the definitions of `Z` and `S`. If we want de/refunctionalization to be a transformation that maps valid programs to valid programs, then the evidence for the productivity of `Z` and `S` has to be composed of the evidence for the termination of `plus` and `mul`. But it is not at all clear how this can be formally specified at the moment.

### 7.2 Universe Hierarchy

Another reason why our system is inconsistent is that we use one impredicative universe with the Type-in-Type axiom. This axiom is known to make the theory inconsistent [Hurkens 1995]; on the other hand, it vastly simplifies the presentation and implementation of the theory if we don't have to care about universe levels. We want to investigate how de- and refunctionalization interact with the assignment of type universes to types. This was also identified as a problem by Huang and Yallop [2023].

### 7.3 The Variance Problem

Most proof assistants enforce *strict positivity* in the definition of data types. The strict positivity restriction says that recursive occurrences of the type that is defined are only allowed at strictly positive positions, and is required to avoid Curry's paradox.

The only source of contravariance in most systems is the function type, where arguments are contravariant. In a system with user-defined codata types many different types are the source of contravariance, but this is quite simple to specify. The problem is that many useful types from object-oriented programming require both positive and negative occurrences of the type being defined. For example, consider the following type[7]:

```
codata NatSet { member(x: Nat): Bool, union(x: NatSet): NatSet }
```

In this example, the `NatSet` type occurs both positively and negatively in the `union` destructor. But this definition is a sensible one; it is not an obscure definition at all. So if we want to enable the user to work with such definitions we have to replace the strict positivity check by something more refined. One avenue that we want to explore is guarded type theory (e.g. [Clouston et al. 2017]). Guarded logic was introduced by [Nakano 2000] precisely in order to fix problems with binary methods in object-oriented programming, and was later developed by other authors into guarded type theories.

### 7.4 Strong Behavioural Equality

There is no universally satisfactory definition of equality as the appropriate definition depends on the object being modeled. For many data types, syntactic equality is sensible. For instance, two natural numbers $n_0, n_1 : \mathbb{N}$ are considered judgmentally equal if they are built from the same constructors, i.e. $\mathtt{Z} \equiv \mathtt{Z}$ and $n_0 \equiv n_1 \implies \mathtt{S}(n_0) \equiv \mathtt{S}(n_1)$. However, the situation is very different as soon as we consider functions $f, g : \mathbb{N} \to \mathbb{N}$. Syntactic equality of the function definitions does not seem appropriate, because multiple definitions can define the *same* function. A more reasonable approach is to regard two functions as equal if they behave identically on all inputs. This principle is known as functional extensionality:

$$\mathtt{fun_ext} : \forall f\ g, (\forall x, \mathtt{Eq}(\mathbb{N} \to \mathbb{N}, f\ x, g\ x)) \implies \mathtt{Eq}(\mathbb{N} \to \mathbb{N}, f, g)$$

Functional extensionality is the prototypical example of behavioral equality. It is a special case of *bisimilarity*: We consider any two objects equal if they behave identically with regard to their observations. For codata types, we often desire behavioral equalities such as bisimilarity.

[7]This example was pointed out in the answer by Neel Krishnaswami to the following question on the proof assistants stack exchange: https://proofassistants.stackexchange.com/questions/372/bringing-oop-features-into-proof-assistants.

In most proof assistants, one can pose those propositional behavioral equalities as axioms. But this approach does not work in our system. This is because the functional extensionality axiom is inconsistent for the data representation Fun of functions $\mathbb{N} \to \mathbb{N}$, which can be seen in the following example:

```
data Fun { Id1, Id2 }
def Fun.ap(x: Nat): Nat { Id1 => x, Id2 => x }
codef apply_id1_eq_id2: Π(Nat, \x. Eq(Fun, Id1.ap(x), Id2.ap(x))) {
    pi_elim(_, _, x) => Refl(Fun, x)
}
fun_ext(Id1, Id2, apply_id1_eq_id2): Eq(Fun, Id1, Id2)
```

Here we can derive the propositional equality Eq(Fun, Id1, Id2), which is a contradiction to the provable proposition that Id1 is not propositionally equal to Id2. Hence, this counterexample shows that we cannot formulate behavioral equalities as axioms in our system.

However, stating behavioral equalities as axioms is often considered unsatisfactory. Axioms break canonicity, i.e., the property that any closed term can be reduced to a canonical value. Further, bisimilarity is always relative to a given set of observations. To see this, consider the functional extensionality axiom from above. It assumes that the function type has a single elimination form, namely function application. Identifying all types that behave identically on application is, without further conditions, only reasonable if application is the only observation we can make. Hence, we have the principal problem that it is unclear how to specify behavioral equalities given that we are extensible in the observations.

Luckily, there are better approaches for working with behavioral equality that are expected to be compatible with our system. For instance, we can resort to the typical Setoid approach of defining a type together with an equality relation:

```
codata Setoid {
    type: Type,
    (self: Setoid).equality: Fun(self.type, Fun(self.type, Type)) }
```

Unfortunately, the Setoid approach is known for bad usability. To achieve a more ergonomic solution, we need a system that allows us to define a type in conjunction with its equalities. In particular, it must be possible for distinctly named constructors to be equal. Future work in this direction could look into extending de- and refunctionalization to observational type theory [Altenkirch and McBride 2006] or a system with higher inductive types.

## 8 RELATED WORK

*Codata types.* Codata types were first introduced by Hagino [Hagino 1987, 1989]. The original interpretation of codata types stems from coalgebras in category theory. An overview of the history of codata types as coalgebras is given by Setzer [Setzer 2012]. We have discussed the relation between codata types and OOP in Section 1.1.

*The expression problem.* The expression problem poses the challenge for statically typed languages to create a type which can be extended by both new producers and new consumers. Based on earlier observations, Wadler [1998] formulated the problem and gave it its current name. The expression problem for proofs is recognized as an important challenge in the verification community, but there are fewer proposed and implemented solutions than in the programming world. One popular solution in the programming world is Swierstra [2008]'s "Data Types à la carte" approach. In that approach, a type is defined as the fixpoint of a coproduct of functors which can be extended by new functors in a modular way. Most proposed solutions for dependent types are based on Swierstra [2008]'s approach and extend them to dependent types. Delaware et al. [2013a,b]; Delaware [2013] as well as Keuchel and Schrijvers [2013] implemented this approach for the Coq proof assistant and Schwaab and Siek [2013] implemented it for Agda. A system for writing modular proofs in the Isabelle proof assistant has been described by Molitor [2015]. The most recent adaptation of the idea is by Forster and Stark [2020], who give an excellent presentation of this line of work in their related work section.

In this article, our focus was not to propose a *solution* to the expression problem for proofs, since the defunctionalization and refunctionalization algorithms are whole-program transformations. Instead, our approach *distills the essence* of the expression problem for dependent types: Neither the functional nor the object-oriented decomposition solves the expression problem, since data types cannot be easily extended by new constructors, and codata types cannot be easily extended by new destructors.

*Dependently-typed object-oriented programming.* In Section 2 we presented our perspective on dependently-typed object-oriented programming. But we are not the first to think about this design space. Jacobs [1995] proposes using coalgebras to express object-oriented classes with coalgebraic specifications. His concept is based on three main components: objects, class implementations, and class specifications. The latter are used to specify a set of methods on an abstract state space as well as a set of assertions that define the behavior of these methods. Such a specification can then be implemented by a class. A class gives a carrier set as a concrete interpretation for the state space and a coalgebra that implements the specified methods. An object is then just an element of the carrier set. In our system, we can express specifications similar to Jacobs' proposal using self-parameters on destructors (see Section 2.2). Rather than having separate notions of specifications, classes, and objects, our system has a singular notion of codata types. Jacobs separates these notions to construct a model in which objects are indistinguishable if they are bisimilar according to their specification. In contrast, in our system, we have a full syntactic duality between data and codata types through de- and refunctionalization. Hence, we need to decouple codata types from the semantics that are usually associated with them, including behavioral equality such as bisimilarity. Setzer [2006] conceived of dependently-typed object-oriented programming by specifying interfaces and having *interactive programs* as objects implementing these interfaces. The interfaces contain a *command type*, which represents the method signatures of an interface. Interactive programs are programs that react to incoming method calls by producing a return value and a new object.

*Dependent type theories with definable* $\Pi$*-type and* $\Sigma$*-type.* In Section 2.3 we demonstrated that the programmer can define both the $\Pi$-type and the $\Sigma$-type in our system, whereas in most proof assistants only the $\Sigma$-type can be defined. This is a generalization of the observation that programmers can't define the function type in most functional programming languages, but that the function type can be defined in object-oriented languages [Setzer 2003]. Apart from this paper, the only other dependent type theory that doesn't presuppose a built-in $\Pi$-type is by Basold [2018]; Basold and Geuvers [2016]. Like us, they give an explicit definition of the $\Pi$-type in their system. Their definition, however, is slightly different from ours, since they have a more expressive core system. In their system, parameterized type constructors and type variables don't have to be fully applied. Partially applied type constructors have a special type $\Gamma \multimap *$, and they specify a sort of simply-typed lambda calculus which governs the rules for abstracting over, and partially applying type constructors to arguments. As a result, their definition of the $\Pi$ and $\Sigma$-type is a bit simpler: We have to use a previously-defined non-dependent function type $A \rightarrow \text{Type}$ to represent the type family that the $\Sigma$ and $\Pi$-types are indexed over, while they use a partially applied type variable $X : A \multimap *$. Our system is also not consistent; theirs is, and they prove both subject reduction and strong normalization.

*Dependent pattern matching.* The traditional primitive elimination forms in dependent type theories are *eliminators*. The eliminator for natural numbers, for example, has the type $\forall (\mathrm{P} : \mathbb{N} \to \mathrm{Type}), \mathrm{P}\ 0 \to (\forall \mathrm{n} : \mathbb{N}, \mathrm{P}\ \mathrm{n} \to \mathrm{P}\ (\mathrm{S}\ \mathrm{n})) \to \forall \mathrm{n} : \mathbb{N}, \mathrm{P}\ \mathrm{n}$. They are suitable for studying the metatheory of dependent types, but programming with them isn't very ergonomic. A more convenient alternative to eliminators is dependent pattern matching, a generalization of ordinary pattern matching to dependent types, which was first proposed by Coquand [1992]. While ordinary pattern matching can be compiled to eliminators and is therefore nothing more than syntactic sugar, the compilation of dependent pattern matches additionally requires Streicher [1993]'s axiom K. Hofmann and Streicher [1994] proved that this axiom does not follow from the standard elimination rules for the identity type. Since the K axiom is sometimes undesirable—it is incompatible with other principles such as univalence—a variant of dependent pattern matching which does not rely on axiom K was developed by Cockx et al. [2014]. We use a variant of dependent pattern and copattern matching which requires axiom K if we want to compile it to eliminators, but we could get rid of this dependency by applying the three restrictions presented in Cockx' thesis [Cockx 2017, p.55].

*Copattern matching.* Copattern matching as a dual concept to pattern matching was first proposed by Abel et al. [2013]. Their work was motivated by the deficiencies of previous approaches which used constructors to represent infinite objects. For instance, the coinductive types originally introduced in Coq broke subject reduction, as noted by Giménez [1996] and Oury [2008]. Even simple infinite objects such as streams cannot be represented using constructors and pattern matching in a sensible way. This follows from the observation of Berger and Setzer [2018] that there exists no decidable equality for streams which admits a one-step expansion of a stream $s$ to a stream (cons $n\ s'$).

*Inconsistent dependent type theories.* The type theory presented in this paper is inconsistent, i.e. every type is inhabited by some term, a property it shares with most programming languages but not with proof assistants. However, the inconsistency of the theory does not imply that the properties expressed by the dependent types are meaningless. We can compare the situation to the programming language Haskell, where it is already possible to write dependent programs by using several language extensions and programming tricks [Eisenberg and Weirich 2012; Lindley and McBride 2013]. Instead of relying on these tricks, a more ergonomic and complete design of dependent types in Haskell has been the subject of various articles [Weirich et al. 2019, 2017] and PhD theses [Eisenberg 2016; Gundry 2013]. Their main insight also applies to our system: the central property of an inconsistent dependent type theory is type soundness [Wright and Felleisen 1994]. For example, every term of type `Vec(5)` can only evaluate to a vector containing five elements or diverge; it cannot evaluate to a vector of six elements. But they also show that inconsistency has downsides, especially for optimization: In a consistent theory every term of type $\mathtt{Eq}(s, t)$ must evaluate to the term `refl`, and can therefore be erased during compilation. In an inconsistent theory, we cannot erase the equality witness, since we could otherwise write a terminating unsafe coercion between arbitrary types, which would violate type soundness.

*Defunctionalization and refunctionalization.* The related work on defunctionalization and refunctionalization can be partitioned into two groups: The first group only considers defunctionalization and refunctionalization for the function type, while the second group generalizes them to transformations between arbitrary data and codata types. De/Refunctionalization of the function type has a long history, which starts with the seminal paper by Reynolds [1972] and the later work of Danvy and Millikin [2009]; Danvy and Nielsen [2001]. That the defunctionalization of polymorphic functions requires GADTs was first observed by Pottier and Gauthier [2006]. In a recent paper, Huang and Yallop [2023] describe the defunctionalization of dependent functions, and especially how to correctly deal with type universes and positivity restrictions, but don't consider the general case of indexed data and codata types. On the contrary, they do not use data types at all and instead introduce the construct of first-class function labels which enables them to avoid problems arising from the expressivity of data type definitions like recursive types. The generalization of defunctionalization from functions to arbitrary codata types was first described by Rendel et al. [2015] for a simply typed system without local lambda abstractions or local pattern matches. That the generalization to polymorphic data and codata types then also requires GAcoDTs has been described by Ostermann and Jabs [2018]. How to treat local pattern and copattern matches in such a way as to preserve the invertibility of defunctionalization and refunctionalization has been described by Binder et al. [2019]. Recently, Zhang et al. [2022] implemented defunctionalization and refunctionalization for the programming language Scala, and used these transformations for some larger case studies. In this article, we describe the generalization to indexed data and codata types, but in distinction to Huang and Yallop [2023] we circumvent the problems of type universes and positivity restrictions by working in an inconsistent type theory.

## 9 CONCLUSION

Most dependently typed programming languages don't support programming with codata as well as programming with data. The main reason some proof assistants support codata types at all was that some support was necessary for the convenient formalization of theorems about infinite and coalgebraic objects. But codata types are useful for more than just representing infinite objects like streams; they represent an orthogonal way to structure programs and proofs, with different extensibility properties and reasoning principles. In this paper we have presented a vision of how programming can look in a dependently typed language in which the data and codata sides are completely symmetric and treated with equal care. By implementing this language and testing it on a case study we have demonstrated that this style of purely functional, dependently typed object-oriented programming does work. We think that this way of systematic language design, in place of ad-hoc extensions, provides a good case study on how the design of dependently typed languages and proof assistants should be approached.

## DATA-AVAILABILITY STATEMENT

This article is accompanied by an online IDE available at polarity-lang.github.io/oopsla24 where the examples discussed in this paper can be selected and loaded. This online IDE consists of a static website hosted on GitHub pages, with all the code running in the browser on the client side. Should the hosted website despite our best efforts no longer be available, then it is possible to recreate it locally using the archived version available at Zenodo [Binder et al. 2024].

# A PROOFS

## A.1 Soundness

In this section, we show that the system introduced in Section 5 is sound with respect to the call-by-value evaluation semantics defined in Section 5.4. We start by introducing some general lemmas before we prove type preservation in Appendix A.2 and progress in Appendix A.3.

First, we are going to prove the inversion lemma. This lemma says that the syntactic form of $e$ determines the last rule in a derivation of $\Gamma \vdash_\Theta e : t$. However, in dependently typed systems the last rule could always have been the following conversion rule:

$$\frac{\Gamma \vdash_\Theta e : t \qquad \Gamma \vdash_\Theta t \equiv t' : \text{Type}}{\Gamma \vdash_\Theta e : t'}\ \textsc{Conv}$$

For this reason we state the inversion lemma in the following more general form:

Lemma A.1 (Inversion). *The typing derivations for producers and consumers can be inverted in the following way.*

- *The last rule in a derivation of* $\Gamma \vdash_\Theta C\sigma : t$ *is either I-Data, I-Codata or Conv.*
- *The last rule in a derivation of* $\Gamma \vdash_\Theta e.d\sigma : t$ *is either E-Data, E-Codata or Conv.*

Proof. By inspection of the rules. □

Lemma A.1 does not put an upper bound on how often the rule Conv occurs in the derivation of a typing judgement. But when we use the inversion lemma we may assume without loss of generality that the rule Conv is used exactly once, since we can always use the reflexivity and transitivity laws for $\equiv$ to bring the derivation into such a normal form. Using this observation, we can prove that every typing derivation of the only possible direct redex $C\overline{v_1}.d\overline{v_2}$ has one of two forms.

Corollary A.2 (Inversion of redexes). *The typing derivation of a redex* $C\overline{v_1}.d\overline{v_2}$ *must end in one of the following two ways. If C is a constructor and d a definition, the typing derivation has the following form:*

$$\textbf{data}\ \mathfrak{T}\Psi\ \{\ C\Xi_1 : \mathfrak{T}\rho_1, \ldots\} \in \Theta$$

$$\cfrac{\cfrac{\vdash_\Theta \overline{v_1} : \Xi_1}{\vdash_\Theta C\overline{v_1} : \mathfrak{T}\rho_1[\overline{v_1}/\Xi_1]}\ \textit{I-Data} \qquad \vdash_\Theta \mathfrak{T}\rho_1[\overline{v_1}/\Xi_1] \equiv \mathfrak{T}\rho_2[\overline{v_2}/\Xi_2]}{\cfrac{\vdash_\Theta C\overline{v_1} : \mathfrak{T}\rho_2[\overline{v_2}/\Xi_2]}{\vdots}}\ \textit{Conv}$$

$$\textbf{def}\ (z : \mathfrak{T}\rho_2).d\Xi_2 : t \in \Theta$$

$$\cfrac{\cfrac{\vdash_\Theta \overline{v_2} : \Xi_2 \qquad \cfrac{\vdots}{\vdash_\Theta C\overline{v_1} : \mathfrak{T}\rho_2[\overline{v_2}/\Xi_2]}\ \textit{Conv}}{\vdash_\Theta C\overline{v_1}.d\overline{v_2} : t[\overline{v_2}/\Xi_2][C\overline{v_1}/z]}\ \textit{E-Data} \qquad \vdash_\Theta t' \equiv t[\overline{v_2}/\Xi_2][C\overline{v_1}/z]}{\vdash_\Theta C\overline{v_1}.d\overline{v_2} : t'}\ \textit{Conv}$$

*Similarly, if C is a codefinition and d is a destructor, the typing derivation ends as follows:*

$$\textbf{codef}\ C\Xi_1 : \mathfrak{T}\rho_1\ \{\ldots\} \in \Theta$$

$$\cfrac{\cfrac{\vdash_\Theta \overline{v_1} : \Xi_1}{\vdash_\Theta C\overline{v_1} : \mathfrak{T}\rho_1[\overline{v_1}/\Xi_1]}\ \textit{I-Codata} \qquad \vdash_\Theta \mathfrak{T}\rho_1[\overline{v_1}/\Xi_1] \equiv \mathfrak{T}\rho_2[\overline{v_2}/\Xi_2]}{\cfrac{\vdash_\Theta C\overline{v_1} : \mathfrak{T}\rho_2[\overline{v_2}/\Xi_2]}{\vdots}}\ \textit{Conv}$$

$$\dfrac{\dfrac{\begin{array}{c}\textbf{codata}\ \mathcal{T}\Psi\ \{ \\ (z : \mathcal{T}\rho_2).d\Xi_2 : t,\ldots\} \in \Theta \\ \vdash_\Theta \overline{v_2} : \Xi_2\end{array} \quad \dfrac{\vdots}{\vdash_\Theta C\overline{v_1} : \mathcal{T}\rho_2[\overline{v_2}/\Xi_2]}\ \textsc{Conv}}{\vdash_\Theta C\overline{v_1}.d\overline{v_2} : t[\overline{v_2}/\Xi_2][C\overline{v_1}/z]}\ \textit{E-}\textsc{Codata} \quad \vdash_\Theta t' \equiv t[\overline{v_2}/\Xi_2][C\overline{v_1}/z]}{\vdash_\Theta C\overline{v_1}.d\overline{v_2} : t'}\ \textsc{Conv}$$

Using Corollary A.2 we can prove that whenever we have a typing derivation for a redex $C\overline{v_1}.d\overline{v_2}$, then the definition d will contain a non-absurd clause for the constructor $C$. (Or similarly, the codefinition C will contain a non-absurd coclause for the destructor d). This lemma is essential for proving progress, since we cannot evaluate to the (non-existing) right-hand side of an absurd clause or coclause.

Lemma A.3 (Existence of non-absurd clause). *If* $\vdash_\Theta C\overline{v_1}.d\overline{v_2} : t'$*, then either*

- $\textbf{def}\ (z : \mathcal{T}\rho).d\Xi_2 : t\ \{C\ \Xi_1 \mapsto e\} \in \Theta$ *or*
- $\textbf{codef}\ C\Xi_1 : \mathcal{T}\rho_1\ \{d\ \Xi_2 \mapsto e\} \in \Theta$

*In particular, we exclude the possibility of an absurd case respectively cocase which does not have a right-hand side.*

Proof. Using Corollary A.2 on the assumption $\vdash_\Theta C\overline{v_1}.d\overline{v_2} : t'$, we get $\vdash_\Theta \mathcal{T}\rho_1[\overline{v_1}/\Xi_1] \equiv \mathcal{T}\rho_2[\overline{v_2}/\Xi_2]$ in both cases. From that, it follows that $\theta = (\overline{v_2}, \overline{v_1})$ unifies $\rho_1$ and $\rho_2$. Therefore, the definition of d is not allowed to ignore the case for constructor C with rule $\textsc{Case}_2$, but must provide a right-hand side with $\textsc{Case}_1$. □

Lemma A.4 (Substitution). *For any program* $\Theta$*, context* $\Gamma$*, expressions* $e$*,* $t$*, and substitution* $\Gamma \vdash \theta : \Xi$*, if* $\Gamma; \Xi \vdash_\Theta e : t$*, then* $\Gamma \vdash_\Theta e[\theta/\Xi] : t[\theta/\Xi]$

Proof. By induction on the derivation of $\Gamma; \Xi \vdash_\Theta e : t$. □

The following lemma expresses that the rules of the system are well-formed in the sense of Martin-Löf. In the special case for judgemental equality this means that whenever we can derive $e_1 \equiv e_2 : t$, then we can also derive that $e_1 : t$ and $e_2 : t$.

Lemma A.5 (Well-formedness). *For all* $e_1, e_2, t$*, if* $\vdash_\Theta e_1 \equiv e_2 : t$*, then* $\vdash_\Theta e_1 : t$ *and* $\vdash_\Theta e_2 : t$*.*

Proof. We prove this by induction on $\vdash_\Theta e_1 \equiv e_2 : t$, where only the cases for $\equiv$-Data and $\equiv$-Codata are interesting. Since these two rules are so similar, we only prove the case for $\equiv$-Data.

$$\dfrac{\begin{array}{cc}\textbf{data}\ \mathcal{T}\Psi\ \{ C\Xi_1 : \mathcal{T}\rho_1,\ldots\} \in \Theta & \vdash_\Theta \sigma_1 : \Xi_1 \\ \textbf{def}\ (z : \mathcal{T}\ \rho_2).d\ \Xi_2 : t\ \{ C\ \Xi_1 \mapsto e,\ldots \} \in \Theta & \vdash_\Theta \sigma_2 : \Xi_2 \\ \multicolumn{2}{c}{\vdash_\Theta \rho_1[\sigma_1/\Xi_1] \equiv \rho_2[\sigma_2/\Xi_2] : \Psi}\end{array}}{\vdash_\Theta C\sigma_1.d\sigma_2 \equiv e[\sigma_2/\Xi_2][\sigma_1/\Xi_1] : t[\sigma_2/\Xi_2][C\sigma_1/z]}\ \equiv\text{-}\textsc{Data}$$

We have to show that both sides of the equality $\equiv$ are well-typed. First, let us show that $C\sigma_1.d\sigma_2$ has type $t[\sigma_2/\Xi_2][C\sigma_1/z]$. For that, we use the following typing derivation. Note that the premise $\vdash_\Theta \mathcal{T}\rho_1[\sigma_1/\Xi_1] \equiv \mathcal{T}\rho_2[\sigma_2/\Xi_2]$ : Type follows from $\vdash_\Theta \rho_1[\sigma_1/\Xi_1] \equiv \rho_2[\sigma_2/\Xi_2] : \Psi$ using the congruence rule for type constructors.

$$\dfrac{\textbf{def}\ (z : \mathcal{T}\rho_2).\mathrm{d}\Xi_2 : t \in \Theta \qquad \vdash_\Theta \sigma_2 : \Xi_2 \qquad \dfrac{\dfrac{\textbf{data}\ \mathcal{T}\Psi\ \{\, \mathrm{C}\Xi_1 : \mathcal{T}\rho_1 \,\} \in \Theta \qquad \vdash_\Theta \sigma_1 : \Xi_1}{\vdash_\Theta \mathrm{C}\sigma_1 : \mathcal{T}\rho_1[\sigma_1/\Xi_1]}\ \textsc{I-Data} \qquad \vdash_\Theta \mathcal{T}\rho_1[\sigma_1/\Xi_1] \equiv \mathcal{T}\rho_2[\sigma_2/\Xi_2]}{\vdash_\Theta \mathrm{C}\sigma_1 : \mathcal{T}\rho_2[\sigma_2/\Xi_2]}\ \textsc{Conv}}{\vdash_\Theta \mathrm{C}\sigma_1.\mathrm{d}\sigma_2 : t[\sigma_2/\Xi_2][\mathrm{C}\sigma_1/\mathrm{z}]}\ \textsc{E-Data}$$

Second, let us show that the right-hand side $e[\sigma_2/\Xi_2][\sigma_1/\Xi_1]$ has type $t[\sigma_2/\Xi_2][\mathrm{C}\sigma_1/\mathrm{z}]$. From the well-formedness of the program, we know the rule $\textsc{Case}_1$ has been used to typecheck the clause for $e$ in the definition **def** $(\mathrm{z} : \mathcal{T}\ \rho_2).\mathrm{d}\ \Xi_2 : t\ \{\ \mathrm{C}\ \Xi_1 \mapsto e,\ldots\ \}$. The rule $\textsc{Case}_1$ has the following form:

$$\dfrac{\textbf{data}\ \mathcal{T}\Psi\ \{\, \mathrm{C}\Xi_1 : \mathcal{T}\rho_1,\ldots\} \in \Theta \qquad \Xi_2;\Xi_1 \vdash_\Theta \theta \text{ mgu for } \rho_1 \equiv \rho_2 : \Psi \qquad (\Xi_2;\Xi_1)[\theta] \vdash_\Theta e[\theta] : t[\mathrm{C}\,\mathrm{id}_{\Xi_1}/\mathrm{z}][\theta]}{\Xi_2 \vdash_\Theta \mathrm{C}\ \Xi_1 \mapsto e : (\mathrm{z} : \mathcal{T}\rho_2).t}\ \textsc{Case}_1$$

From the third premise of that rule we have $(\Xi_2;\Xi_1)[\theta] \vdash_\Theta e[\theta] : t[\mathrm{C}\,\mathrm{id}_{\Xi_1}/\mathrm{z}][\theta]$. Now we need to make use of some properties of the most general unifier $\theta$ of $\rho_1$ and $\rho_2$. From the derivation of $\vdash_\Theta \rho_1[\sigma_1/\Xi_1] \equiv \rho_2[\sigma_2/\Xi_2] : \Psi$ we also know that $(\sigma_1, \sigma_2)$ is a unifier for $\rho_1$ and $\rho_2$. From the property of being a most general unifier, we can deduce that there exists some $\theta'$ such that $(\sigma_1, \sigma_2) = \theta' \circ \theta$. By Lemma A.4 we know that the following judgment holds:

$$(\Xi_2;\Xi_1)[\theta] \vdash e[\theta] : t[\mathrm{C}\,\mathrm{id}_{\Xi_1}/\mathrm{z}][\theta] \tag{1}$$

$$\vdash e[\theta][\theta'] : t[\mathrm{C}\,\mathrm{id}_{\Xi_1}/\mathrm{z}][\theta][\theta'] \tag{2}$$

$$\vdash e[(\sigma_1, \sigma_2)] : t[\mathrm{C}\,\mathrm{id}_{\Xi_1}/\mathrm{z}][(\sigma_1, \sigma_2)] \tag{3}$$

$$\vdash e[\sigma_2/\Xi_2][\sigma_1/\Xi_1] : t[\sigma_2/\Xi_2][\mathrm{C}\sigma_1/\mathrm{z}] \tag{4}$$

The derivability of (2) follows from the derivability of (1) by the use of the substitution Lemma A.4, since $\theta' : \emptyset \mapsto (\Xi_2, \Xi_1)[\theta]$. The derivability of (3) follows by the observation from above that $(\sigma_1, \sigma_2)$ is defined as $\theta' \circ \theta$, which corresponds to first applying the substitution $\theta$ and then applying the substitution $\theta'$. Finally, the derivability of (4) follows by standard properties of substitution and using the fact that $t$ only contains free variables from $\Xi_2$,

□

## A.2 Type Preservation

We use evaluation contexts in our formalization of evaluation. The first lemma shows that if an evaluation context containing a redex is typeable, then the redex itself is also typable.

Lemma A.6 (Extract typing under evaluation contexts). *If* $\Gamma \vdash_\Theta E[e] : t$, *then there exists a* $t'$ *such that* $\Gamma \vdash_\Theta e : t'$.

Proof. By induction on the typing derivation $\Gamma \vdash_\Theta E[e] : t$. Note that the context $\Gamma$ is the same for $E[e]$ and $e$, since no variables are bound in the evaluation context $E$. This reflects that we only evaluate to weak normal form, i.e. we don't evaluate under binders. □

Lemma A.7 (Equality under evaluation contexts). *If* $\Gamma \vdash_\Theta e : t$ *and* $\Gamma \vdash_\Theta E[e] : t'$, *then for all* $e'$ *for which* $\Gamma \vdash_\Theta e \equiv e' : t$, *we have* $\Gamma \vdash_\Theta E[e'] : t'$.

Proof. By induction on the structure of $E$ and by applying the appropriate congruence rules for each term construct. □

Thanks to the previous lemmas, we only have to show the preservation theorem for direct redexes.

Lemma A.8 (Evaluation implies equality). *If* $\vdash_\Theta e_1 : t$ *and* $e_1 \triangleright_\beta e_2$*, then* $\vdash_\Theta e_1 \equiv e_2 : t$

Proof. Since $\triangleright_\beta$ only applies to direct redexes, we know that $e_1$ must have the form $\mathrm{C}\overline{v_1}.\mathrm{d}\overline{v_2}$. From the typing derivation $\vdash_\Theta \mathrm{C}\overline{v_1}.\mathrm{d}\overline{v_2} : t$, Lemma A.3, and Corollary A.2 we obtain all the necessary premises for the rule ≡-Data resp. ≡-Codata:

$$\frac{\begin{array}{ll}\textbf{data}\ \mathcal{T}\Psi\ \{\ \mathrm{C}\Xi_1 : \mathcal{T}\rho_1, \ldots\} \in \Theta & \vdash_\Theta \sigma_1 : \Xi_1 \\ \textbf{def}\ (\mathrm{z} : \mathcal{T}\ \rho_2).\mathrm{d}\ \Xi_2 : t\ \{\ \mathrm{C}\ \Xi_1 \mapsto e, \ldots\ \} \in \Theta & \vdash_\Theta \sigma_2 : \Xi_2 \\ \multicolumn{2}{c}{\vdash_\Theta \rho_1[\sigma_1/\Xi_1] \equiv \rho_2[\sigma_2/\Xi_2] : \Psi}\end{array}}{\vdash_\Theta \mathrm{C}\sigma_1.\mathrm{d}\sigma_2 \equiv e[\sigma_2/\Xi_2][\sigma_1/\Xi_1] : t[\sigma_2/\Xi_2][\mathrm{C}\sigma_1/\mathrm{z}]}\ \equiv\text{-Data}$$

□

Theorem 5.2 (Preservation). *For any well-formed program* $\Theta$ *and any expressions* $e_1, e_2, t$ *if* $\vdash_\Theta e_1 : t$ *and* $e_1 \triangleright e_2$*, then* $\vdash_\Theta e_2 : t$.

Proof. From the way we have defined reduction in Section 5.4, we know that the evaluation has the following form:

$$e_1 = E[e_1'] \triangleright E[e_2'] = e_2 \qquad e_1' \triangleright_\beta e_2'$$

From Lemma A.6, we know that $e_1'$ is typeable for some $t'$. From Lemma A.8, we know that $\vdash_\Theta e_1' \equiv e_2' : t'$. From Lemma A.5, we know that $\vdash_\Theta e_2' : t'$. Finally, by Lemma A.7, we get $\vdash_\Theta e_2 : t$ which concludes the proof. □

## A.3 Progress

The progress theorem says that for any closed expression which is not yet a value we can make progress by taking one evaluation step. In order to prove this theorem we first have to introduce some smaller lemmas. The first lemma states that any closed expression which is not a value can be uniquely decomposed into an evaluation context and a redex.

Lemma A.9 (Unique Decomposition). *For any closed expression* $e$*, either* $e$ *is a value or there exist unique* $E, \overline{v_1}, \overline{v_2}$ *such that* $e = E[\mathrm{C}\overline{v_1}.\mathrm{d}\overline{v_2}]$

Proof. By induction on $e$ and the grammar for evaluation contexts and values. Note that the expression must be closed, otherwise we would have to consider non-values involving free variables. □

With these lemmas in place, we can now prove the main progress theorem.

Theorem 5.1 (Progress). *For any well-formed program* $\Theta$ *and expressions* $e$ *and* $t$*, if* $\vdash_\Theta e : t$ *then either* $e$ *is a value or there exists an expression* $e'$ *such that* $e \triangleright e'$.

Proof. By Lemma A.9, $e$ must either be a value, in which case we are finished, or it must be of the form $E[\mathrm{C}\overline{v_1}.\mathrm{d}\overline{v_2}]$. By Lemma A.6, we know that there exists a $t'$ such that $\vdash_\Theta \mathrm{C}\overline{v_1}.\mathrm{d}\overline{v_2} : t'$. By Lemma A.3, we know that the corresponding case in the (co)pattern match is non-absurd. One of the following applies:

**Data** d is a definition **def** $(z : \mathcal{T}\rho).\mathrm{d}\Xi_2 : t \{\mathrm{C}\ \Xi_1 \mapsto s\} \in \Theta$. Then we can apply the following evaluation rule:

$$\frac{\textbf{def}\ (z : \mathcal{T}\rho).\mathrm{d}\Xi_2 : t\ \{\mathrm{C}\ \Xi_1 \mapsto s\} \in \Theta}{\mathrm{C}\overline{v_1}.\mathrm{d}\overline{v_2} \triangleright_\beta s[\overline{v_2}/\Xi_2][\overline{v_1}/\Xi_1]}$$

Let $e' := E[s[\overline{v_1}/\Xi_1][\overline{v_2}/\Xi_2]]$ and then the proposition follows by congruence.

**Codata** C is a codefinition **codef** $\mathrm{C}\Xi_1 : \mathcal{T}\rho_1\ \{\mathrm{d}\ \Xi_2 \mapsto s\} \in \Theta$. Then we can apply the following evaluation rule:

$$\frac{\textbf{codef}\ \mathrm{C}\Xi_1 : \mathcal{T}\rho_1\ \{\mathrm{d}\ \Xi_2 \mapsto s\} \in \Theta}{\mathrm{C}\overline{v_1}.\mathrm{d}\overline{v_2} \triangleright_\beta s[\overline{v_1}/\Xi_1][\overline{v_2}/\Xi_2]}$$

Let $e' := E[s[\overline{v_1}/\Xi_1][\overline{v_2}/\Xi_2]]$ and then the proposition follows by congruence.

□

This concludes the proof of type soundness for the language.

### A.4 De- and Refunctionalization Preserve Typing

In this section, we sketch the proofs for the propositions stated in Section 6. First, we want to show Theorem 6.3, which states that de- and refunctionalization preserve typing and judgmental equality. We separately prove the result for de- and refunctionalization in Lemmas A.10 and A.11.

Lemma A.10 (Defunctionalization preserves typing and judgmental equality). *The following implications hold:*

$$\begin{array}{llll}
\mathcal{H}_1 & \Gamma \vdash_\Theta e : t & \Longrightarrow & \Gamma \vdash_{\mathcal{D}_\mathcal{T}(\Theta)} e : t \\
\mathcal{H}_2 & \Gamma \vdash_\Theta e_1 \equiv e_2 : t & \Longrightarrow & \Gamma \vdash_{\mathcal{D}_\mathcal{T}(\Theta)} e_1 \equiv e_2 : t \\
\mathcal{H}_3 & \Gamma \vdash_\Theta \sigma : \Xi & \Longrightarrow & \Gamma \vdash_{\mathcal{D}_\mathcal{T}(\Theta)} \sigma : \Xi \\
\mathcal{H}_4 & \Gamma \vdash_\Theta \sigma_1 \equiv \sigma_2 : \Xi & \Longrightarrow & \Gamma \vdash_{\mathcal{D}_\mathcal{T}(\Theta)} \sigma_1 \equiv \sigma_2 : \Xi \\
\mathcal{H}_5 & \vdash_\Theta \Gamma\ \mathrm{ctx} & \Longrightarrow & \vdash_{\mathcal{D}_\mathcal{T}(\Theta)} \Gamma\ \mathrm{ctx} \\
\mathcal{H}_6 & \Gamma \vdash_\Theta \Xi\ \mathrm{tel} & \Longrightarrow & \Gamma \vdash_{\mathcal{D}_\mathcal{T}(\Theta)} \Xi\ \mathrm{tel}
\end{array}$$

Let us briefly discuss the proof idea. Recall that expressions are not syntactically affected by defunctionalization. This is because we reuse the syntax for *producers* for both constructor and codefinition calls and the syntax for *consumers* for both destructor and definition calls (see Figure 9). The same applies to substitutions, contexts, and telescopes. It remains to show that we can construct a typing derivation given the defunctionalized program $\mathcal{D}_\mathcal{T}(\Theta)$. Of course, this is only interesting if the original typing derivation involves the defunctionalized type $\mathcal{T}$. All top-level declarations other than $\mathcal{T}$ remain unchanged in $\mathcal{D}_\mathcal{T}(\Theta)$.

Proof. By simultaneous induction. We show each of the implications $\mathcal{H}_1 \ldots \mathcal{H}_6$ by case distinction on the inference rule used for the root of the derivation. In this proof sketch, we limit ourselves to show $\mathcal{H}_2$ for the computation rule ≡-Codata. Assume ≡-Codata was used to derive the following judgmental equality:

$$\frac{\vdots}{\Gamma \vdash_\Theta \mathrm{C}\ \sigma_1.\mathrm{d}\ \sigma_2 \equiv e[\sigma_1/\Xi_1][\sigma_2/\Xi_2] : t[\sigma_2/\Xi_2][\mathrm{C}\ \sigma_1/\mathrm{z}]}\ (\equiv\text{-Codata})$$

We need to show that this equation still holds after defunctionalizing $\mathcal{T}$ in $\Theta$, i.e.

$$\Gamma \vdash_{\mathcal{D}_\mathcal{T}(\Theta)} \mathrm{C}\ \sigma_1.\mathrm{d}\ \sigma_2 \equiv e[\sigma_1/\Xi_1][\sigma_2/\Xi_2] : t[\sigma_2/\Xi_2][\mathrm{C}\ \sigma_1/\mathrm{z}]$$

We consider the case where C is a codefinition of the type $\mathcal{T}$ being defunctionalized. Then we get the following assumptions from the premises of ≡-Codata:

$\mathcal{A}_1$: $\textbf{codata}\ \mathcal{T}\ \Psi\ \{\ \ldots, (\mathrm{z} : \mathcal{T}\rho_2).\ \mathrm{d}\ \Xi_2 : t, \ldots\ \} \in \Theta$
$\mathcal{A}_2$: $\textbf{codef}\ \mathrm{C}\ \Xi_1 : \mathcal{T}\ \rho_1\ \{\ \mathrm{d}\ \Xi_2 \mapsto e, \ldots\ \} \in \Theta$
$\mathcal{A}_3$: $\Gamma \vdash_\Theta \sigma_1 : \Xi_1$
$\mathcal{A}_4$: $\Gamma \vdash_\Theta \sigma_2 : \Xi_2$
$\mathcal{A}_5$: $\Gamma \vdash_\Theta \rho_1[\sigma_1/\Xi_1] \equiv \rho_2[\sigma_2/\Xi_2] : \Psi$

The original definition of $\mathcal{T}$ in $\Theta$ reads as follows:

$$\textbf{codata}\ \mathcal{T}\ \Psi\ \{\ (\mathrm{z} : \mathcal{T}\rho_2).\ \mathrm{d}\ \Xi_2 : t, \ldots\ \}, \textbf{codef}\ \mathrm{C}\ \Xi_1 : \mathcal{T}\ \rho_1\ \{\ \mathrm{d}\ \Xi_2 \mapsto e, \ldots\ \}$$

Defunctionalization of $\mathcal{T}$ turns the codefinition C into a constructor on $\mathcal{T}$. The destructor d becomes a top-level definition. The transformed declaration of $\mathcal{T}$ within $\mathcal{D}_{\mathcal{T}}(\Theta)$ is:

$$\textbf{data}\ \mathcal{T}\ \Psi\ \{\ \mathrm{C}\ \Xi_1 : \mathcal{T}\ \rho_1, \ldots\ \}, \textbf{def}\ (\mathrm{z} : \mathcal{T}\ \rho_2).\mathrm{d}\ \Xi_2 : t\ \{\ \mathrm{C}\ \Xi_1 \mapsto e, \ldots\ \}$$

Hence, we can show the goal by applying ≡-Data as follows:

$$\dfrac{\begin{array}{ll} \textbf{data}\ \mathcal{T}\ \Psi\ \{\ \ldots, \mathrm{C}\ \Xi_1 : \mathcal{T}\ \rho_1, \ldots\ \} \in \mathcal{D}_{\mathcal{T}}(\Theta) & \text{by def. of } \mathcal{D}_{\mathcal{T}}(\Theta) \\ \textbf{def}\ (\mathrm{z} : \mathcal{T}\ \rho_2).\mathrm{d}\ \Xi_2 : t\ \{\ \mathrm{C}\ \Xi_1 \mapsto e, \ldots\ \} \in \mathcal{D}_{\mathcal{T}}(\Theta) & \text{by def. of } \mathcal{D}_{\mathcal{T}}(\Theta) \\ \Gamma \vdash_{\mathcal{D}_{\mathcal{T}}(\Theta)} \sigma_1 : \Xi_1 & \text{by IH } \mathcal{H}_3 \text{ on } \mathcal{A}_3 \\ \Gamma \vdash_{\mathcal{D}_{\mathcal{T}}(\Theta)} \sigma_2 : \Xi_2 & \text{by IH } \mathcal{H}_3 \text{ on } \mathcal{A}_4 \\ \Gamma \vdash_{\mathcal{D}_{\mathcal{T}}(\Theta)} \rho_1[\sigma_1/\Xi_1] \equiv \rho_2[\sigma_2/\Xi_2] : \Psi & \text{by IH } \mathcal{H}_4 \text{ on } \mathcal{A}_5 \end{array}}{\Gamma \vdash_{\mathcal{D}_{\mathcal{T}}(\Theta)} \mathrm{C}\ \sigma_1.\mathrm{d}\ \sigma_2 \equiv e[\sigma_1/\Xi_1][\sigma_2/\Xi_2] : t[\sigma_2/\Xi_2][\mathrm{C}\ \sigma_1/\mathrm{z}]}\ (\equiv\text{-DATA})$$

All premises follow by definition of the defunctionalized program $\Theta$ or by applying the induction hypothesis.

Let us now briefly discuss the rest of the proof. Again, we only consider the interesting cases where the original derivation concerns the defunctionalized type $\mathcal{T}$. The remaining cases follow by applying the induction hypotheses.

- $\mathcal{H}_1$: Here we need to consider type formation, introduction, and elimination. Upon reviewing Figure 10, it should be easy to see the correspondence between the rules for data and codata types. For example, if the original derivation uses I-Codata, we need to apply I-Data in the defunctionalized program. The reasoning is similar to the case shown above.
- $\mathcal{H}_2$: We have shown the case for ≡-Codata above. By similar reasoning, the result follows for the typical congruence rules for judgmental equality.
- $\mathcal{H}_3 \ldots \mathcal{H}_6$: Substitutions and telescopes are only affected by defunctionalization if the typing of subexpressions changes. The inference rule used to construct the root of the derivation remains the same. Hence, these cases follow by applying the induction hypotheses where needed.

□

Lemma A.11 (Refunctionalization preserves typing and judgmental equality). *The following implications hold:*

| | | | |
|---|---|---|---|
| $\mathcal{H}_1$ | $\Gamma \vdash_\Theta e : t$ | $\Longrightarrow$ | $\Gamma \vdash_{\mathcal{R}_{\mathcal{T}}(\Theta)} e : t$ |
| $\mathcal{H}_2$ | $\Gamma \vdash_\Theta e_1 \equiv e_2 : t$ | $\Longrightarrow$ | $\Gamma \vdash_{\mathcal{R}_{\mathcal{T}}(\Theta)} e_1 \equiv e_2 : t$ |
| $\mathcal{H}_3$ | $\Gamma \vdash_\Theta \sigma : \Xi$ | $\Longrightarrow$ | $\Gamma \vdash_{\mathcal{R}_{\mathcal{T}}(\Theta)} \sigma : \Xi$ |
| $\mathcal{H}_4$ | $\Gamma \vdash_\Theta \sigma_1 \equiv \sigma_2 : \Xi$ | $\Longrightarrow$ | $\Gamma \vdash_{\mathcal{R}_{\mathcal{T}}(\Theta)} \sigma_1 \equiv \sigma_2 : \Xi$ |
| $\mathcal{H}_5$ | $\vdash_\Theta \Gamma\ \mathrm{ctx}$ | $\Longrightarrow$ | $\vdash_{\mathcal{R}_{\mathcal{T}}(\Theta)} \Gamma\ \mathrm{ctx}$ |
| $\mathcal{H}_6$ | $\Gamma \vdash_\Theta \Xi\ \mathrm{tel}$ | $\Longrightarrow$ | $\Gamma \vdash_{\mathcal{R}_{\mathcal{T}}(\Theta)} \Xi\ \mathrm{tel}$ |

Proof. By simultaneous induction. The proof is entirely symmetric to the proof of Lemma A.10.

□

Theorem 6.3 (De/Refunctionalization preserves typing and judgmental equality). *The following implications hold:*

- $\Gamma \vdash_\Theta e : t \implies \Gamma \vdash_{\mathcal{X}_\mathcal{T}(\Theta)} e : t$
- $\Gamma \vdash_\Theta e_1 \equiv e_2 : t \implies \Gamma \vdash_{\mathcal{X}_\mathcal{T}(\Theta)} e_1 \equiv e_2 : t$
- $\Gamma \vdash_\Theta \sigma : \Xi \implies \Gamma \vdash_{\mathcal{X}_\mathcal{T}(\Theta)} \sigma : \Xi$
- $\Gamma \vdash_\Theta \sigma_1 \equiv \sigma_2 : \Xi \implies \Gamma \vdash_{\mathcal{X}_\mathcal{T}(\Theta)} \sigma_1 \equiv \sigma_2 : \Xi$
- $\vdash_\Theta \Gamma \text{ ctx} \implies \vdash_{\mathcal{X}_\mathcal{T}(\Theta)} \Gamma \text{ ctx}$
- $\Gamma \vdash_\Theta \Xi \text{ tel} \implies \Gamma \vdash_{\mathcal{X}_\mathcal{T}(\Theta)} \Xi \text{ tel}$

Proof. Direct corollary of Lemmas A.10 and A.11. □

Theorem 6.4 (De/Refunctionalization preserves well-formedness of programs). *If* $\vdash_\Theta \Theta$ *Ok, then* $\vdash_\Theta \mathcal{X}_\mathcal{T}(\Theta)$ *Ok*

Proof. Assuming that all top-level definitions in the original program $\Theta$ are well-typed, we need to show that the program $\mathcal{D}_\mathcal{T}(\Theta)$, where $\mathcal{T}$ has been de- or refunctionalized is also well-typed. First, de- and refunctionalization only affect the (co)data type $\mathcal{T}$ and all (co)definitions defined for it. Hence, it suffices to show that the transformed (co)data type and the transformed (co)definitions are well-typed. In this proof sketch, we consider the refunctionalization of a data type $\mathcal{T}$. To improve the readability, we only explicitly write a single constructor C and a single definition d for $\mathcal{T}$ to represent the definitions for $\mathcal{T}$ in $\Theta$. Hence, in the original program $\Theta$ we have:

$$\textbf{data}\ \mathcal{T}\ \Psi\ \{\ \mathrm{C}\ \Xi_1 : \mathcal{T}\ \rho_1, \dots\ \}, \textbf{def}\ (\mathrm{z} : \mathcal{T}\ \rho_2).\mathrm{d}\ \Xi_2 : t\ \{\ \mathrm{C}\ \Xi_1 \mapsto e, \dots\ \}$$

Refunctionalization of $\mathcal{T}$ yields $\mathcal{R}_\mathcal{T}(\Theta)$:

$$\textbf{codata}\ \mathcal{T}\ \Psi\ \{\ (\mathrm{z} : \mathcal{T}\rho_2).\ \mathrm{d}\ \Xi_2 : t, \dots\ \}, \textbf{codef}\ \mathrm{C}\ \Xi_1 : \mathcal{T}\ \rho_1\ \{\ \mathrm{d}\ \Xi_2 \mapsto e, \dots\ \}$$

Our assumption is that $\Theta$ is well-typed, hence we get the following assumptions from the derivation of $\vdash \Theta$ Ok:

$$\frac{\begin{array}{l}\mathcal{A}_1 :\ \vdash_\Theta \Psi \text{ tel} \\ \mathcal{A}_2 :\ \vdash_\Theta \Xi_1 \text{ tel} \\ \mathcal{A}_3 :\ \Xi_1 \vdash_\Theta \rho_1 : \Psi \\ \dots\end{array}}{\vdash_\Theta \textbf{data}\ \mathcal{T}\ \Psi\ \{\ \mathrm{C}\ \Xi_1 : \mathcal{T}\ \rho_1, \dots\ \}\ \text{Ok}}\ (\text{Data})$$

$$\frac{\begin{array}{l}\mathcal{B}_1 :\ \textbf{data}\ \mathcal{T}\ \Psi\ \{\ \dots\ \} \in \Theta \\ \mathcal{B}_2 :\ \Xi_2 \vdash_\Theta \rho_2 : \Psi \\ \mathcal{B}_3 :\ \Xi_2; \mathrm{z} : \mathcal{T}\ \rho_2 \vdash_\Theta t : \text{Type} \\ \mathcal{B}_4 :\ \Xi_2 \vdash_\Theta \mathrm{C}\ \Xi_1 \mapsto e : (\mathrm{z} : \mathcal{T}\ \rho_2).t, \dots\end{array}}{\vdash_\Theta \textbf{def}\ (\mathrm{z} : \mathcal{T}\ \rho_2).\mathrm{d}\ \Xi_2 : t\ \{\ \dots\ \}\ \text{Ok}}\ (\text{Def})$$

$$\frac{\begin{array}{l}C_1 :\ \textbf{data}\ \mathcal{T}\ \Psi\ \{\ \dots, \mathrm{C}\ \Xi_1 : \mathcal{T}\ \rho_2, \dots\ \} \in \Theta \\ C_2 :\ \Xi_2; \Xi_1 \vdash_\Theta \theta \text{ mgu for } \rho_1 \equiv \rho_2 : \Psi \\ C_3 :\ (\Xi_2; \Xi_1)[\theta] \vdash_\Theta e[\theta] : t[\mathrm{C}\ \mathrm{id}_{\Xi_1}/\mathrm{z}][\theta]\end{array}}{\Xi_2 \vdash_\Theta \mathrm{C}\ \Xi_1 \mapsto e : (\mathrm{z} : \mathcal{T}\ \rho_2).t}\ (\text{Case}_1)$$

We can now type the refunctionalized program $\Theta_{\mathcal{R}} := \mathcal{R}_{\mathcal{T}}(\Theta)$ as follows. We start by giving the typing derivation for the new codata type $\mathcal{T}$:

$$\dfrac{\begin{array}{lr} \vdash_{\Theta_{\mathcal{R}}} \Psi \text{ tel} & \text{by Lemma A.11 on } \mathcal{A}_1 \\ \vdash_{\Theta_{\mathcal{R}}} \Xi_2 \text{ tel} & \text{by Lemma A.11 on presupposition of } \mathcal{B}_2 \\ \Xi_2 \vdash_{\Theta_{\mathcal{R}}} \rho_2 : \Psi & \text{by Lemma A.11 on } \mathcal{B}_2 \\ \Xi_2, z : \mathcal{T}\ \rho_2 \vdash_{\Theta_{\mathcal{R}}} t : \text{Type} & \text{by Lemma A.11 on } \mathcal{B}_3 \end{array}}{\vdash_{\Theta_{\mathcal{R}}} \textbf{codata}\ \mathcal{T}\ \Psi\ \{\ \overline{(\mathrm{z} : \mathcal{T}\rho_2).\ \mathrm{d}\ \Xi_2 : t}\ \}\ \textsc{Ok}}\ (\textsc{Codata})$$

We can type the codefinition C as follows:

$$\dfrac{\begin{array}{lr} \textbf{codata}\ \mathcal{T}\ \Psi\ \{\ \dots\ \} \in \Theta_{\mathcal{R}} & \text{by def. of } \Theta_{\mathcal{R}} \\ \Xi_1 \vdash_{\Theta_{\mathcal{R}}} \rho_1 : \Psi & \text{by Lemma A.11 on } \mathcal{A}_3 \\ \Xi_1 \vdash_{\Theta_{\mathcal{R}}} \mathrm{d}\ \Xi_2 \mapsto e : (\mathrm{C} : \mathcal{T}\ \rho_1), \dots & \text{by the derivation below} \end{array}}{\vdash_{\Theta_{\mathcal{R}}} \textbf{codef}\ \mathrm{C}\ \Xi_1 : \mathcal{T}\ \rho_1\ \{\ \mathrm{d}\ \Xi_2 \mapsto e, \dots\ \}\ \textsc{Ok}}\ (\textsc{Codef})$$

where we type cocases as follows:

$$\dfrac{\begin{array}{lr} \textbf{codata}\ \mathcal{T}\ \Psi\ \{\ (\mathrm{z} : \mathcal{T}\rho_2).\ \mathrm{d}\ \Xi_2 : t\ \} \in \Theta_{\mathcal{R}} & \text{by def. of } \Theta_{\mathcal{R}} \\ \Xi_1; \Xi_2 \vdash_{\Theta_{\mathcal{R}}} \theta \text{ mgu for } \rho_1 \equiv \rho_2 : \Psi & \text{by } \mathcal{C}_2,\ \Xi_1 \# \Xi_2 \\ (\Xi_1; \Xi_2)[\theta] \vdash_{\Theta_{\mathcal{R}}} e[\theta] : t[\mathrm{C}\ \mathrm{id}_{\Xi_1}/\mathrm{z}][\theta] & \text{by Lemma A.11 on } \mathcal{C}_3,\ \Xi_1 \# \Xi_2 \end{array}}{\Xi_1 \vdash_{\Theta_{\mathcal{R}}} \mathrm{d}\ \Xi_2 \mapsto e : (\mathrm{C} : \mathcal{T}\ \rho_1)}\ (\textsc{Cocase}_1)$$

Most premises follow by definition of the refunctionalized program $\Theta_{\mathcal{R}}$ or by applying Lemma A.11. For the last two premises in the derivation using $\textsc{Cocase}_1$, we need to exchange the order of $\Xi_1$ and $\Xi_2$. This works because these telescopes are defined independently in the program. By applying $\alpha$-renaming as needed, we also ensure that no shadowing takes place, i.e., $\Xi_1 \# \Xi_2$.

□

# B WEB SERVER CASE STUDY: COMPLETE SOURCE CODE

This appendix contains the complete source code for the case study in Section 3.

## B.1 Before Refunctionalization

The complete code for the case study before refunctionalization is shown below:

```
data Nat { Z, S(n: Nat) }
data Bool { True, False }
codata Fun(a b: Type) {
   Fun(a, b).ap(a b: Type, x: a): b }
codata Π(A: Type, T: A -> Type) {
   Π(A, T).dap(A: Type, T: A -> Type, x: A): T.ap(A, Type, x) }
codata ×₋(a b: Type) {
   ×₋(a, b).fst(a b: Type): a,
   ×₋(a, b).snd(a b: Type): b }
codef Pair(a b: Type, x: a, y: b): ×₋(a, b) {
   fst(a, b) => x,
   snd(a, b) => y }
data Response { Forbidden, Return(n: Nat) }
codata User { hasCredentials: Bool }
codata State(loggedIn: Bool) {
   State(False).login(u: User): State(u.hasCredentials),
   State(True).logout: State(False),
   State(True).increment: State(True),
   State(True).set(n: Nat): State(True),
   (self: State(True)).set_idempotent(b: Bool, n: Nat)
      : Eq(State(True), self.set(n), self.set(n).set(n)),
   (self: State(True)).setResult(b: Bool, n: Nat): Eq(Nat, n, self.set(n).counter(True)),
   State(b).counter(b: Bool): Nat }
codata Utils {
```

```
   put_twice(n: Nat, route: Route, state: State(route.requiresLogin))
      : ×₋(State(route.requiresLogin), Response) }
codef MkUtils: Utils {
   put_twice(n, route, state) =>
      route.put(n)
           .ap(State(route.requiresLogin),
               ×₋(State(route.requiresLogin), Response),
               route.put(n)
                    .ap(State(route.requiresLogin), ×₋(State(route.requiresLogin), Response), state)
                    .fst(State(route.requiresLogin), Response)) }
data Eq(t: Type, a b: t) {
   Refl(t: Type, a: t): Eq(t, a, a) }
data Route { Index }
def (self: Route).put_idempotent(n: Nat)
   : Π(State(self.requiresLogin),
       \state. Eq(×₋(State(self.requiresLogin), Response),
                  self.put(n)
                      .ap(State(self.requiresLogin),
                          ×₋(State(self.requiresLogin), Response),
                          state),
                  MkUtils.put_twice(n, self, state))) {
   Index =>
      comatch {
         dap(_, _, state) =>
            Refl(×₋(State(False), Response), Pair(State(False), Response, state, Forbidden)) } }
def (self: Route).post: State(self.requiresLogin) -> ×₋(State(self.requiresLogin), Response) {
   Index =>
      \state. comatch {
         fst(a, b) => state,
         snd(a, b) => Forbidden } }
def Route.requiresLogin: Bool { Index => False }
def (self: Route).get: State(self.requiresLogin) -> Response {
   Index => \state. Return(state.counter(False)) }
def (self: Route).put(n: Nat)
   : State(self.requiresLogin) -> ×₋(State(self.requiresLogin), Response) {
   Index => \state. Pair(State(False), Response, state, Forbidden) }
```

## B.2 After Refunctionalizing and Adding New Route

The complete code for the case study after refunctionalization and adding the `Admin` route is shown below:

```
data Nat { Z, S(n: Nat) }
data Bool { True, False }
codata Fun(a b: Type) {
   Fun(a, b).ap(a b: Type, x: a): b }
codata Π(A: Type, T: A -> Type) {
   Π(A, T).dap(A: Type, T: A -> Type, x: A): T.ap(A, Type, x) }
codata ×₋(a b: Type) {
   ×₋(a, b).fst(a b: Type): a,
   ×₋(a, b).snd(a b: Type): b }
codef Pair(a b: Type, x: a, y: b): ×₋(a, b) {
   fst(a, b) => x,
   snd(a, b) => y }
data Response { Forbidden, Return(n: Nat) }
codata User { hasCredentials: Bool }
codata State(loggedIn: Bool) {
   State(False).login(u: User): State(u.hasCredentials),
   State(True).logout: State(False),
   State(True).increment: State(True),
   State(True).set(n: Nat): State(True),
   (self: State(True)).set_idempotent(b: Bool, n: Nat)
      : Eq(State(True), self.set(n), self.set(n).set(n)),
   (self: State(True)).setResult(b: Bool, n: Nat): Eq(Nat, n, self.set(n).counter(True)),
   State(b).counter(b: Bool): Nat }
codata Utils {
   put_twice(n: Nat, route: Route, state: State(route.requiresLogin))
      : ×₋(State(route.requiresLogin), Response) }
codef MkUtils: Utils {
   put_twice(n, route, state) =>
      route.put(n)
```

```
          .ap(State(route.requiresLogin),
              ×₋(State(route.requiresLogin), Response),
              route.put(n)
                   .ap(State(route.requiresLogin), ×₋(State(route.requiresLogin), Response), state)
                   .fst(State(route.requiresLogin), Response)) }
data Eq(t: Type, a b: t) {
   Refl(t: Type, a: t): Eq(t, a, a) }
def Eq(t1, a, b).cong_pair(t1 t2: Type, a b: t1, c: t2)
   : Eq(×₋(t1, t2), Pair(t1, t2, a, c), Pair(t1, t2, b, c)) {
   Refl(_, _) => Refl(×₋(t1, t2), Pair(t1, t2, b, c)) }
codata Route {
   requiresLogin: Bool,
   (self: Route).get: State(self.requiresLogin) -> Response,
   (self: Route).post: State(self.requiresLogin) -> ×₋(State(self.requiresLogin), Response),
   (self: Route).put(n: Nat): State(self.requiresLogin) -> ×₋(State(self.requiresLogin), Response),
   (self: Route).put_idempotent(n: Nat)
      : Π(State(self.requiresLogin),
          \state. Eq(×₋(State(self.requiresLogin), Response),
                     self.put(n)
                         .ap(State(self.requiresLogin),
                             ×₋(State(self.requiresLogin), Response),
                             state),
                     MkUtils.put_twice(n, self, state))) }
codef Index: Route {
   requiresLogin => False,
   post =>
      \state. comatch {
         fst(a, b) => state,
         snd(a, b) => Forbidden },
   get => \state. Return(state.counter(False)),
   put(n) => \state. Pair(State(False), Response, state, Forbidden),
   put_idempotent(n) =>
      comatch {
         dap(_, _, state) =>
            Refl(×₋(State(False), Response), Pair(State(False), Response, state, Forbidden)) } }
codef Admin: Route {
   requiresLogin => True,
   post =>
      \state. comatch {
         fst(a, b) => state.increment,
         snd(a, b) => Return(state.increment.counter(True)) },
   get => \state. Return(state.counter(True)),
   put(n) => \state. Pair(State(True), Response, state.set(n), Return(n)),
   put_idempotent(n) =>
      comatch {
         dap(_, _, state) =>
            state.set_idempotent(True, n)
                 .cong_pair(State(True), Response, state.set(n), state.set(n).set(n), Return(n)) } }
```

## C DE- AND REFUNCTIONALIZATION OF SOUNDNESS PROOFS

In the main part of the paper we studied how the duality between data and codata types influences how we can write dependently typed programs. We focused on programming instead of proofs because the language we described is just sound, but not consistent. In this appendix, we explore how an object-oriented style of proving might look like in a consistent version of the language. For this reason we formalize a preservation proof for a small expression language in Appendix C.1. We observe that extending this proof with new evaluation rules is difficult, we therefore switch to the object-oriented view in Appendix C.2 and show how we can simply extend the proof in that decomposition. In order to demonstrate that even more complex proofs can be expressed in our language, we provide a soundness proof for the simply-typed lambda calculus in Appendix C.3.

### C.1 A Preservation Proof in a Functional Proof Assistant

The simple expression language that we formalize has three term constructors: The two boolean values and an if-then-else expression. There is also only one type, the type of Booleans.

```
-- | The terms of the object language.
data Tm { TmTrue, TmFalse, TmIte(c e1 e2: Tm) }
-- | The types of the object language.
data Ty { TyBool }
```

We formalize the operational semantics of the language in the small-step style. There are two rules, one rule for evaluating an if-then-else applied to the `TmTrue` value, and one congruence rule for evaluating the condition of the if-then-else expression.

```
-- | Small step operational semantics.
data Step(e1 e2: Tm) {
   StIteT(e1 e2: Tm): Step(TmIte(TmTrue, e1, e2), e1),
   StIte(e1 e2 e3 e4: Tm, s: Step(e1, e2)): Step(TmIte(e1, e3, e4), TmIte(e2, e3, e4)) }
```

There is one typing rule for each term of the language: Boolean values can be typed as a Boolean, and an if-then-else requires the condition to be a boolean and both branches to have the same type.

```
-- | The typing relation.
data Typing(e: Tm, ty: Ty) {
   TTrue: Typing(TmTrue, TyBool),
   TFalse: Typing(TmFalse, TyBool),
   TIte(e1 e2 e3: Tm, ty: Ty, t1: Typing(e1, TyBool), t2: Typing(e2, ty), t3: Typing(e3, ty))
      : Typing(TmIte(e1, e2, e3), ty) }
```

This is enough to formalize and prove the preservation property for this simple expression language. The complete state of the proof looks like this:

```
-- | The terms of the object language.
data Tm { TmTrue, TmFalse, TmIte(c e1 e2: Tm) }
-- | The types of the object language.
data Ty { TyBool }
-- | Small step operational semantics.
data Step(e1 e2: Tm) {
   StIteT(e1 e2: Tm): Step(TmIte(TmTrue, e1, e2), e1),
   StIte(e1 e2 e3 e4: Tm, s: Step(e1, e2)): Step(TmIte(e1, e3, e4), TmIte(e2, e3, e4)) }
-- | The typing relation.
data Typing(e: Tm, ty: Ty) {
   TTrue: Typing(TmTrue, TyBool),
   TFalse: Typing(TmFalse, TyBool),
   TIte(e1 e2 e3: Tm, ty: Ty, t1: Typing(e1, TyBool), t2: Typing(e2, ty), t3: Typing(e3, ty))
      : Typing(TmIte(e1, e2, e3), ty) }
-- | Preservation.
codata Preservation(e: Tm, ty: Ty) {
   Preservation(e1, ty).preservationStep(e1 e2: Tm, ty: Ty, s: Step(e1, e2)): Typing(e2, ty) }
def Typing(e, ty).pres(e: Tm, ty: Ty): Preservation(e, ty) {
   TTrue =>
      comatch PreservationTrue {
         preservationStep(e1, e2, ty, s) =>
            s.match {
               StIteT(_, _) absurd,
               StIte(_, _, _, _, _) absurd } },
   TFalse =>
      comatch PreservationFalse {
         preservationStep(e1, e2, ty, s) =>
            s.match {
               StIteT(_, _) absurd,
               StIte(_, _, _, _, _) absurd } },
   TIte(e1, e2, e3, ty, t1, t2, t3) =>
      comatch PreservationIte {
         preservationStep(e4, e5, ty, s) =>
            s.match {
               StIteT(_, _) => t2,
               StIte(e4, e6, e7, e8, s) =>
                  TIte(e6, e7, e8, ty,
                       t1.pres(e4, TyBool).preservationStep(e4, e6, TyBool, s),
                       t2, t3) } } }
```

The alert reader might already have realized that we forgot a sensible reduction rule: We did not specify what happens when we encounter an if-then-else applied to the value `TmFalse`. In the next section, we will address this issue.

## C.2 Extending the Proof in an Object-Oriented Proof Assistant

We will now try to find the best way to add the missing reduction rule. In the current state of our program, adding the corresponding rule would require us to add a new constructor to the Step data type. We would then have to search the whole (admittedly quite small) program for pattern matches on the Step data type, so that we could add the missing case for our new constructor. Using refunctionalization allows us to try a different approach: First, we refunctionalize the Step data type. This produces a new program where Step is a codata type, its old constructors are now codefinitions and the matches in Typing are replaced by calls to the destructors of the new Step codata type. Note that each destructor of the Step codata type corresponds to one of the pattern matches on Step in the original program.

```
-- | Small step operational semantics.
codata Step(e1 e2: Tm) {
   Step(TmFalse, e2).d_step3(e2: Tm): Typing(e2, TyBool),
   Step(TmTrue, e2).d_step1(e2: Tm): Typing(e2, TyBool),
   Step(TmIte(e1, e2, e3), e5).d_step5(ty: Ty,
                                       e1 e2 e3: Tm,
                                       t1: Typing(e1, TyBool),
                                       t2: Typing(e2, ty),
                                       t3: Typing(e3, ty),
                                       e5: Tm )
      : Typing(e5, ty) }
codef StIteT(e1 e2: Tm): Step(TmIte(TmTrue, e1, e2), e1) {
   d_step3(e3) absurd,
   d_step1(e3) absurd,
   d_step5(e3, e4, e5, e5, ty, t2, t3) => t2 }
codef StIte(e1 e2 e3 e4: Tm, s: Step(e1, e2)): Step(TmIte(e1, e3, e4), TmIte(e2, e3, e4)) {
   d_step5(ty, e5, e6, e7, t1, t2, t3, e8) =>
      TIte(e2, e6, e7, ty, t1.pres(e5, TyBool).preservationStep(e5, e2, TyBool, s), t2, t3),
   d_step3(e5) absurd,
   d_step1(e5) absurd }
def Typing(e, ty).pres(e: Tm, ty: Ty): Preservation(e, ty) {
   TTrue => comatch PreservationTrue { preservationStep(e1, e2, ty0, s) => s.d_step1(e2) },
   TFalse => comatch PreservationFalse { preservationStep(e1, e2, ty0, s) => s.d_step3(e2) },
   TIte(e1, e2, e3, ty0, t1, t2, t3) =>
      comatch PreservationIte {
         preservationStep(e', e'', ty1, s) => s.d_step5(e1, e2, e3, e'', ty, t2, t3) } }
```

Now adding the missing rule simply requires us to write one new codefinition which has one cocase for every match on Step that occurred in the original program.

```
-- | The terms of the object language.
data Tm { TmTrue, TmFalse, TmIte(c e1 e2: Tm) }
-- | The types of the object language.
data Ty { TyBool }
-- | Small step operational semantics.
codata Step(e1 e2: Tm) {
   Step(TmFalse, e2).d_step3(e2: Tm): Typing(e2, TyBool),
   Step(TmTrue, e2).d_step1(e2: Tm): Typing(e2, TyBool),
   Step(TmIte(e1, e2, e3), e5).d_step5(e1 e2 e3 e5: Tm,
                                       ty: Ty,
                                       t1: Typing(e1, TyBool),
                                       t2: Typing(e2, ty),
                                       t3: Typing(e3, ty) )
      : Typing(e5, ty) }
codef StIteT(e1 e2: Tm): Step(TmIte(TmTrue, e1, e2), e1) {
   d_step3(e3) absurd,
   d_step1(e3) absurd,
   d_step5(e3, e4, e5, e5, ty, t1, t2, t3) => t2 }
codef StIteF(e1 e2: Tm): Step(TmIte(TmFalse, e1, e2), e2) {
   d_step3(e3) absurd,
   d_step1(e3) absurd,
   d_step5(e3, e4, e5, e5, ty, t1, t2, t3) => t3 }
codef StIte(e1 e2 e3 e4: Tm, s: Step(e1, e2)): Step(TmIte(e1, e3, e4), TmIte(e2, e3, e4)) {
   d_step1(e5) absurd,
   d_step3(e5) absurd,
   d_step5(e1', e2', e3', e5', ty, t1, t2, t3) =>
```

```
    TIte(e2, e3, e4, ty, t1.pres(e1, TyBool).preservationStep(e1, e2, TyBool, s), t2, t3) }
-- | The typing relation.
data Typing(e: Tm, ty: Ty) {
  TTrue: Typing(TmTrue, TyBool),
  TFalse: Typing(TmFalse, TyBool),
  TIte(e1 e2 e3: Tm, ty: Ty, t1: Typing(e1, TyBool), t2: Typing(e2, ty), t3: Typing(e3, ty))
    : Typing(TmIte(e1, e2, e3), ty) }
-- | Preservation.
codata Preservation(e: Tm, ty: Ty) {
  Preservation(e1, ty).preservationStep(e1 e2: Tm, ty: Ty, s: Step(e1, e2)): Typing(e2, ty) }
def Typing(e, ty).pres(e: Tm, ty: Ty): Preservation(e, ty) {
  TTrue => comatch PreservationTrue { preservationStep(e1, e2, ty0, s) => s.d_step1(e2) },
  TFalse => comatch PreservationFalse { preservationStep(e1, e2, ty0, s) => s.d_step3(e2) },
  TIte(e1, e2, e3, ty0, t1, t2, t3) =>
    comatch PreservationIte {
      preservationStep(e4, e5, ty1, s) => s.d_step5(e1, e2, e3, e5, ty, t1, t2, t3) } }
```

As a final step, we could defunctionalize `Step` again, which would result in a program similar to the original, but with a new constructor `StIteF` for `Step` and an additional case for this new constructor in each pattern match of this type. However, this is not strictly necessary: The program works fine as it is and thus this step is only really required when we want to add further pattern matches on `Step`.

In this example we have shown how to add a missing rule to the operational semantics, but similar problems of modularity can appear for any of the types which are involved. We might want to add a new term to the expressions of the language, or a new type. We might want to add a new typing rule or an additional proof. Therefore, it may be beneficial for proof assistants to support both paradigms. For the simple development in this section, the problem is not yet really urgent, so we have provided an implementation of a type soundness proof for the simply typed lambda calculus in Appendix C.3. Our implementation allows us to interactively explore how the proof changes if any of the involved types are changed from the data to the codata representation, or vice versa.

### C.3 Type Soundness of the Simply-Typed Lambda Calculus

In the previous subsection we introduced a small case study for an expression language without variables and binders. In this section we extend this to a full soundness proof of the simply-typed lambda calculus.

```
data Exp { Var(x: Nat), Lam(body: Exp), App(lhs rhs: Exp) }
data Typ { FunT(t1 t2: Typ), VarT(x: Nat) }
data Ctx { Nil, Cons(t: Typ, ts: Ctx) }
def Ctx.append(other: Ctx): Ctx {
  Nil => other,
  Cons(t, ts) => Cons(t, ts.append(other)) }
def Ctx.len: Nat {
  Nil => Z,
  Cons(_, ts) => S(ts.len) }
def Exp.subst(v: Nat, by: Exp): Exp {
  Var(x) => x.cmp(v).subst_result(x, by),
  Lam(e) => Lam(e.subst(S(v), by)),
  App(e1, e2) => App(e1.subst(v, by), e2.subst(v, by)) }
def Cmp.subst_result(x: Nat, by: Exp): Exp {
  LT => Var(x),
  EQ => by,
  GT => Var(x.pred) }
data Elem(x: Nat, t: Typ, ctx: Ctx) {
  Here(t: Typ, ts: Ctx): Elem(Z, t, Cons(t, ts)),
  There(x: Nat, t t2: Typ, ts: Ctx, prf: Elem(x, t, ts)): Elem(S(x), t, Cons(t2, ts)) }
data HasType(ctx: Ctx, e: Exp, t: Typ) {
  TVar(ctx: Ctx, x: Nat, t: Typ, elem: Elem(x, t, ctx)): HasType(ctx, Var(x), t),
  TLam(ctx: Ctx, t1 t2: Typ, e: Exp, body: HasType(Cons(t1, ctx), e, t2))
    : HasType(ctx, Lam(e), FunT(t1, t2)),
  TApp(ctx: Ctx,
      t1 t2: Typ,
```

```
        e1 e2: Exp,
        e1_t: HasType(ctx, e1, FunT(t1, t2)),
        e2_t: HasType(ctx, e2, t1) )
      : HasType(ctx, App(e1, e2), t2) }
data Eval(e1 e2: Exp) {
   EBeta(e1 e2: Exp): Eval(App(Lam(e1), e2), e1.subst(Z, e2)),
   ECongApp1(e1 e1': Exp, h: Eval(e1, e1'), e2: Exp): Eval(App(e1, e2), App(e1', e2)),
   ECongApp2(e1 e2 e2': Exp, h: Eval(e2, e2')): Eval(App(e1, e2), App(e1, e2')) }
data IsValue(e: Exp) {
   VLam(e: Exp): IsValue(Lam(e)) }
data Progress(e: Exp) {
   PVal(e: Exp, h: IsValue(e)): Progress(e),
   PStep(e1 e2: Exp, h: Eval(e1, e2)): Progress(e1) }
def (e: Exp).progress(t: Typ): HasType(Nil, e, t) -> Progress(e) {
   Var(x) =>
      \h_t. h_t.match {
         TVar(_, _, _, elem) => elem.empty_absurd(x, t).elim_bot(Progress(Var(x))),
         TLam(_, _, _, _, _) absurd,
         TApp(_, _, _, _, _, _, _) absurd },
   Lam(e) => \_. PVal(Lam(e), VLam(e)),
   App(e1, e2) =>
      \h_t. h_t.match {
         TVar(_, _, _, _) absurd,
         TLam(_, _, _, _, _) absurd,
         TApp(_, t1, t2, _, _, e1_t, e2_t) =>
            e1.progress(FunT(t1, t2)).ap(HasType(Nil, e1, FunT(t1, t2)), Progress(e1), e1_t).match {
               PStep(_, e1', e1_eval_e1') =>
                  PStep(App(e1, e2), App(e1', e2), ECongApp1(e1, e1', e1_eval_e1', e2)),
               PVal(_, is_val) =>
                  is_val.match { VLam(e) => PStep(App(Lam(e), e2), e.subst(Z, e2), EBeta(e, e2)) } } } }
def (e1: Exp).preservation(e2: Exp, t: Typ)
   : HasType(Nil, e1, t) -> Eval(e1, e2) -> HasType(Nil, e2, t) {
   Var(_) =>
      \h_t. \h_eval. h_eval.match {
         EBeta(_, _) absurd,
         ECongApp1(_, _, _, _) absurd,
         ECongApp2(_, _, _, _) absurd },
   Lam(_) =>
      \h_t. \h_eval. h_eval.match {
         EBeta(_, _) absurd,
         ECongApp1(_, _, _, _) absurd,
         ECongApp2(_, _, _, _) absurd },
   App(e1, e2) =>
      \h_t. h_t.match {
         TVar(_, _, _, _) absurd,
         TLam(_, _, _, _, _) absurd,
         TApp(_, t1, t2, _, _, h_lam, h_e2) =>
            \h_eval. h_eval.match {
               ECongApp1(_, e1', h, _) =>
                  TApp(Nil, t1, t2,
                       e1', e2,
                       e1.preservation(e1', FunT(t1, t2))
                         .ap(HasType(Nil, e1, FunT(t1, t2)),
                             Eval(e1, e1') -> HasType(Nil, e1', FunT(t1, t2)),
                             h_lam)
                         .ap(Eval(e1, e1'), HasType(Nil, e1', FunT(t1, t2)), h),
                       h_e2),
               ECongApp2(_, _, e2', h) =>
                  TApp(Nil, t1, t2, e1,
                       e2',
                       h_lam,
                       e2.preservation(e2', t1)
                         .ap(HasType(Nil, e2, t1), Eval(e2, e2') -> HasType(Nil, e2', t1), h_e2)
                         .ap(Eval(e2, e2'), HasType(Nil, e2', t1), h)),
               EBeta(e1, _) =>
                  h_lam.match {
                     TVar(_, _, _, _) absurd,
                     TApp(_, _, _, _, _, _, _) absurd,
                     TLam(_, _, _, _, h_e1) =>
                        e1.subst_lemma(Nil, Nil, t1, t2, e2)
                          .ap(HasType(Cons(t1, Nil), e1, t2),
```

```
                                HasType(Nil, e2, t1) -> HasType(Nil, e1.subst(Z, e2), t2),
                                h_e1)
                             .ap(HasType(Nil, e2, t1), HasType(Nil, e1.subst(Z, e2), t2), h_e2) } } } }
def (e: Exp).subst_lemma(ctx1 ctx2: Ctx, t1 t2: Typ, by_e: Exp)
  : HasType(ctx1.append(Cons(t1, ctx2)), e, t2) -> HasType(Nil,
                                                           by_e,
                                                           t1) -> HasType(ctx1.append(ctx2),
                                                                      e.subst(ctx1.len, by_e),
                                                                      t2) {
  Var(x) =>
    \h_e. \h_by. h_e.match {
      TLam(_, _, _, _, _) absurd,
      TApp(_, _, _, _, _, _, _) absurd,
      TVar(_, _, _, h_elem) =>
        x.cmp_reflect(ctx1.len).match {
          IsLT(_, _, h_eq_lt, h_lt) =>
            h_eq_lt.transport(Cmp,
                              LT,
                              x.cmp(ctx1.len),
                              comatch {
                                ap(_, _, cmp) =>
                                  HasType(ctx1.append(ctx2), cmp.subst_result(x, by_e), t2) },
                              ctx2.weaken_append(ctx1, Var(x), t2)
                                  .ap(HasType(ctx1, Var(x), t2),
                                    HasType(ctx1.append(ctx2), Var(x), t2),
                                    TVar(ctx1, x, t2,
                                        ctx1.elem_append_first(Cons(t1, ctx2), t2, x)
                                            .ap(LE(S(x), ctx1.len),
                                               Elem(x, t2,
                                                   ctx1.append(Cons(t1,
                                                                    ctx2))) -> Elem(x, t2,
                                                                                   ctx1),
                                               h_lt)
                                            .ap(Elem(x, t2, ctx1.append(Cons(t1, ctx2))),
                                               Elem(x, t2, ctx1),
                                               h_elem)))),
          IsEQ(_, _, h_eq_eq, h_eq) =>
            h_eq_eq.transport(Cmp,
                              EQ,
                              x.cmp(ctx1.len),
                              comatch {
                                ap(_, _, cmp) =>
                                  HasType(ctx1.append(ctx2), cmp.subst_result(x, by_e), t2) },
                              ctx1.append(ctx2)
                                  .weaken_append(Nil, by_e, t2)
                                  .ap(HasType(Nil, by_e, t2),
                                    HasType(ctx1.append(ctx2), by_e, t2),
                                    ctx1.ctx_lookup(ctx2, t2, t1)
                                        .ap(Elem(ctx1.len, t2, ctx1.append(Cons(t1, ctx2))),
                                          Eq(Typ, t1, t2),
                                          h_eq.transport(Nat,
                                                         x,
                                                         ctx1.len,
                                                         comatch {
                                                           ap(_, _, x) =>
                                                             Elem(x, t2,
                                                                 ctx1.append(Cons(t1,
                                                                                  ctx2))) },
                                                         h_elem))
                                        .transport(Typ,
                                                   t1, t2,
                                                   comatch {
                                                     ap(_, _, t) => HasType(Nil, by_e, t) },
                                                   h_by))),
          IsGT(_, _, h_eq_gt, h_gt) =>
            h_eq_gt.transport(Cmp,
                              GT,
                              x.cmp(ctx1.len),
                              comatch {
                                ap(_, _, cmp) =>
                                  HasType(ctx1.append(ctx2), cmp.subst_result(x, by_e), t2) },
```

```
                                 TVar(ctx1.append(ctx2),
                                      x.pred,
                                      t2,
                                      ctx1.elem_append_pred(ctx2, t2, t1, x)
                                          .ap(LE(S(ctx1.len), x),
                                              Elem(x, t2,
                                                   ctx1.append(Cons(t1, ctx2))) -> Elem(x.pred,
                                                                                        t2,
                                                                                        ctx1.append(ctx2)),
                                              h_gt)
                                          .ap(Elem(x, t2, ctx1.append(Cons(t1, ctx2))),
                                              Elem(x.pred, t2, ctx1.append(ctx2)),
                                              h_elem))) } },
   Lam(body) =>
      \h_e. \h_by. h_e.match {
         TVar(_, _, _, _) absurd,
         TApp(_, _, _, _, _, _, _) absurd,
         TLam(_, a, b, _, h_body) =>
            TLam(ctx1.append(ctx2),
                 a, b,
                 body.subst(S(ctx1.len), by_e),
                 body.subst_lemma(Cons(a, ctx1), ctx2, t1, b, by_e)
                     .ap(HasType(Cons(a, ctx1).append(Cons(t1, ctx2)), body, b),
                         HasType(Nil, by_e, t1) -> HasType(Cons(a, ctx1).append(ctx2),
                                                           body.subst(S(ctx1.len), by_e),
                                                           b),
                         h_body)
                     .ap(HasType(Nil, by_e, t1),
                         HasType(Cons(a, ctx1).append(ctx2), body.subst(S(ctx1.len), by_e), b),
                         h_by)) },
   App(e1, e2) =>
      \h_e. \h_by. h_e.match {
         TVar(_, _, _, _) absurd,
         TLam(_, _, _, _, _) absurd,
         TApp(_, a, b, _, _, h_e1, h_e2) =>
            TApp(ctx1.append(ctx2),
                 a, b,
                 e1.subst(ctx1.len, by_e),
                 e2.subst(ctx1.len, by_e),
                 e1.subst_lemma(ctx1, ctx2, t1, FunT(a, b), by_e)
                   .ap(HasType(ctx1.append(Cons(t1, ctx2)), e1, FunT(a, b)),
                       HasType(Nil, by_e, t1) -> HasType(ctx1.append(ctx2),
                                                         e1.subst(ctx1.len, by_e),
                                                         FunT(a, b)),
                       h_e1)
                   .ap(HasType(Nil, by_e, t1),
                       HasType(ctx1.append(ctx2), e1.subst(ctx1.len, by_e), FunT(a, b)),
                       h_by),
                 e2.subst_lemma(ctx1, ctx2, t1, a, by_e)
                   .ap(HasType(ctx1.append(Cons(t1, ctx2)), e2, a),
                       HasType(Nil, by_e, t1) -> HasType(ctx1.append(ctx2),
                                                         e2.subst(ctx1.len, by_e),
                                                         a),
                       h_e2)
                   .ap(HasType(Nil, by_e, t1),
                       HasType(ctx1.append(ctx2), e2.subst(ctx1.len, by_e), a),
                       h_by)) } }
def (ctx2: Ctx).weaken_append(ctx1: Ctx, e: Exp, t: Typ)
   : HasType(ctx1, e, t) -> HasType(ctx1.append(ctx2), e, t) {
   Nil =>
      \h_e. ctx1.append_nil
               .transport(Ctx,
                          ctx1,
                          ctx1.append(Nil),
                          comatch { ap(_, _, ctx) => HasType(ctx, e, t) },
                          h_e),
   Cons(t', ts) =>
      \h_e. ctx1.append_assoc(Cons(t', Nil), ts)
               .transport(Ctx,
                          ctx1.append(Cons(t', Nil)).append(ts),
                          ctx1.append(Cons(t', ts)),
```

```
                              comatch { ap(_, _, ctx) => HasType(ctx, e, t) },
                              ts.weaken_append(ctx1.append(Cons(t', Nil)), e, t)
                                .ap(HasType(ctx1.append(Cons(t', Nil)), e, t),
                                    HasType(ctx1.append(Cons(t', Nil)).append(ts), e, t),
                                    e.weaken_cons(ctx1, t', t)
                                     .ap(HasType(ctx1, e, t),
                                         HasType(ctx1.append(Cons(t', Nil)), e, t),
                                         h_e))) }
def (e: Exp).weaken_cons(ctx: Ctx, t1 t2: Typ)
   : HasType(ctx, e, t2) -> HasType(ctx.append(Cons(t1, Nil)), e, t2) {
   Var(x) =>
      \h_e. h_e.match {
         TLam(_, _, _, _, _) absurd,
         TApp(_, _, _, _, _, _, _) absurd,
         TVar(_, _, _, h_elem) =>
            TVar(ctx.append(Cons(t1, Nil)), x, t2, h_elem.elem_append(x, t1, t2, ctx)) },
   Lam(e) =>
      \h_e. h_e.match {
         TVar(_, _, _, _) absurd,
         TApp(_, _, _, _, _, _, _) absurd,
         TLam(_, a, b, _, h_e) =>
            TLam(ctx.append(Cons(t1, Nil)),
                 a, b, e,
                 e.weaken_cons(Cons(a, ctx), t1, b)
                  .ap(HasType(Cons(a, ctx), e, b),
                      HasType(Cons(a, ctx).append(Cons(t1, Nil)), e, b),
                      h_e)) },
   App(e1, e2) =>
      \h_e. h_e.match {
         TVar(_, _, _, _) absurd,
         TLam(_, _, _, _, _) absurd,
         TApp(_, a, b, _, _, h_e1, h_e2) =>
            TApp(ctx.append(Cons(t1, Nil)),
                 a, b, e1, e2,
                 e1.weaken_cons(ctx, t1, FunT(a, b))
                   .ap(HasType(ctx, e1, FunT(a, b)),
                       HasType(ctx.append(Cons(t1, Nil)), e1, FunT(a, b)),
                       h_e1),
                 e2.weaken_cons(ctx, t1, a)
                   .ap(HasType(ctx, e2, a), HasType(ctx.append(Cons(t1, Nil)), e2, a), h_e2)) } }
def Elem(n, t2, ctx).elem_append(n: Nat, t1 t2: Typ, ctx: Ctx)
   : Elem(n, t2, ctx.append(Cons(t1, Nil))) {
   Here(t, ts) => Here(t, ts.append(Cons(t1, Nil))),
   There(n, _, t', ts, h) =>
      There(n, t2, t', ts.append(Cons(t1, Nil)), h.elem_append(n, t1, t2, ts)) }
def (ctx1: Ctx).append_assoc(ctx2 ctx3: Ctx)
   : Eq(Ctx, ctx1.append(ctx2).append(ctx3), ctx1.append(ctx2.append(ctx3))) {
   Nil => Refl(Ctx, ctx2.append(ctx3)),
   Cons(x, xs) =>
      xs.append_assoc(ctx2, ctx3)
        .cong(Ctx,
              Ctx,
              xs.append(ctx2).append(ctx3),
              xs.append(ctx2.append(ctx3)),
              comatch { ap(_, _, xs) => Cons(x, xs) }) }
def (ctx: Ctx).append_nil: Eq(Ctx, ctx, ctx.append(Nil)) {
   Nil => Refl(Ctx, Nil),
   Cons(t, ts) => ts.append_nil.eq_cons(ts, ts.append(Nil), t) }
def Elem(x, t, Nil).empty_absurd(x: Nat, t: Typ): Bot {
   Here(_, _) absurd,
   There(_, _, _, _, _) absurd }
def Elem(Z, t1, Cons(t2, ctx)).elem_unique(ctx: Ctx, t1 t2: Typ): Eq(Typ, t2, t1) {
   Here(_, _) => Refl(Typ, t1),
   There(_, _, _, _, _) absurd }
def (ctx1: Ctx).ctx_lookup(ctx2: Ctx, t1 t2: Typ)
   : Elem(ctx1.len, t1, ctx1.append(Cons(t2, ctx2))) -> Eq(Typ, t2, t1) {
   Nil => \h. h.elem_unique(ctx2, t1, t2),
   Cons(t, ts) =>
      \h. h.match {
         Here(_, _) absurd,
         There(_, _, _, _, h) =>
```

```
            ts.ctx_lookup(ctx2, t1, t2)
              .ap(Elem(ts.len, t1, ts.append(Cons(t2, ctx2))), Eq(Typ, t2, t1), h) } }
def (ctx1: Ctx).elem_append_first(ctx2: Ctx, t: Typ, x: Nat)
   : LE(S(x), ctx1.len) -> Elem(x, t, ctx1.append(ctx2)) -> Elem(x, t, ctx1) {
   Nil =>
      \h_lt. \h_elem. h_lt.match {
         LERefl(_) absurd,
         LESucc(_, _, _) absurd },
   Cons(t', ts) =>
      \h_lt. \h_elem. h_elem.match {
         Here(_, _) => Here(t, ts),
         There(x', _, _, _, h) =>
            There(x', t,
                  t', ts,
                  ts.elem_append_first(ctx2, t, x')
                    .ap(LE(S(x'), ts.len),
                        Elem(x', t, ts.append(ctx2)) -> Elem(x', t, ts),
                        h_lt.le_unsucc(x, ts.len))
                    .ap(Elem(x', t, ts.append(ctx2)), Elem(x', t, ts), h)) } }
def (ctx1: Ctx).elem_append_pred(ctx2: Ctx, t1 t2: Typ, x: Nat)
   : LE(S(ctx1.len), x) -> Elem(x, t1, ctx1.append(Cons(t2, ctx2))) -> Elem(x.pred,
                                                                             t1,
                                                                             ctx1.append(ctx2)) {
   Nil =>
      \h_gt. \h_elem. h_elem.match {
         Here(_, _) =>
            h_gt.match {
               LERefl(_) absurd,
               LESucc(_, _, _) absurd },
         There(_, _, _, _, h) => h },
   Cons(t, ts) =>
      \h_gt. \h_elem. h_elem.match {
         Here(_, _) =>
            h_gt.match {
               LERefl(_) absurd,
               LESucc(_, _, _) absurd },
         There(x', _, _, _, h) =>
            h_gt.le_unsucc(S(ts.len), x')
                .s_pred(ts.len, x')
                .transport(Nat,
                           S(x'.pred),
                           x',
                           comatch { ap(_, _, x) => Elem(x, t1, Cons(t, ts).append(ctx2)) },
                           There(x'.pred,
                                 t1, t,
                                 ts.append(ctx2),
                                 ts.elem_append_pred(ctx2, t1, t2, x')
                                   .ap(LE(S(ts.len), x'),
                                       Elem(x', t1, ts.append(Cons(t2, ctx2))) -> Elem(x'.pred,
                                                                                       t1,
                                                                                       ts.append(ctx2)),
                                       h_gt.le_unsucc(S(ts.len), x'))
                                   .ap(Elem(x', t1, ts.append(Cons(t2, ctx2))),
                                       Elem(x'.pred, t1, ts.append(ctx2)),
                                       h))) } }
data Bot { }
def Bot.elim_bot(a: Type): a { }
codata Fun(a b: Type) {
   Fun(a, b).ap(a b: Type, x: a): b }
data Eq(a: Type, x y: a) {
   Refl(a: Type, x: a): Eq(a, x, x) }
def Eq(a, x, y).sym(a: Type, x y: a): Eq(a, y, x) { Refl(a, x) => Refl(a, x) }
def Eq(a, x, y).transport(a: Type, x y: a, p: a -> Type, prf: p.ap(a, Type, x)): p.ap(a, Type, y) {
   Refl(a, x) => prf }
def Eq(a, x, y).cong(a b: Type, x y: a, f: a -> b): Eq(b, f.ap(a, b, x), f.ap(a, b, y)) {
   Refl(a, x) => Refl(b, f.ap(a, b, x)) }
def Eq(Nat, x, y).eq_s(x y: Nat): Eq(Nat, S(x), S(y)) { Refl(_, _) => Refl(Nat, S(x)) }
def Eq(Ctx, xs, ys).eq_cons(xs ys: Ctx, t: Typ): Eq(Ctx, Cons(t, xs), Cons(t, ys)) {
   Refl(_, _) => Refl(Ctx, Cons(t, xs)) }
data Nat { Z, S(n: Nat) }
def Nat.pred: Nat {
```

```
   Z => Z,
   S(x) => x }
data LE(x y: Nat) {
   LERefl(x: Nat): LE(x, x),
   LESucc(x y: Nat, h: LE(x, y)): LE(x, S(y)) }
data Cmp { LT, EQ, GT }
def Nat.cmp(y: Nat): Cmp {
   Z =>
      y.match {
         Z => EQ,
         S(_) => LT },
   S(x) =>
      y.match {
         Z => GT,
         S(y) => x.cmp(y) } }
data CmpReflect(x y: Nat) {
   IsLT(x y: Nat, h1: Eq(Cmp, LT, x.cmp(y)), h2: LE(S(x), y)): CmpReflect(x, y),
   IsEQ(x y: Nat, h1: Eq(Cmp, EQ, x.cmp(y)), h2: Eq(Nat, x, y)): CmpReflect(x, y),
   IsGT(x y: Nat, h1: Eq(Cmp, GT, x.cmp(y)), h2: LE(S(y), x)): CmpReflect(x, y) }
def (x: Nat).cmp_reflect(y: Nat): CmpReflect(x, y) {
   Z =>
      y.match as y => CmpReflect(Z, y) {
         Z => IsEQ(Z, Z, Refl(Cmp, EQ), Refl(Nat, Z)),
         S(y) => IsLT(Z, S(y), Refl(Cmp, LT), y.z_le.le_succ(Z, y)) },
   S(x) =>
      y.match as y => CmpReflect(S(x), y) {
         Z => IsGT(S(x), Z, Refl(Cmp, GT), x.z_le.le_succ(Z, x)),
         S(y) =>
            x.cmp_reflect(y).match {
               IsLT(_, _, h1, h2) => IsLT(S(x), S(y), h1, h2.le_succ(S(x), y)),
               IsEQ(_, _, h1, h2) => IsEQ(S(x), S(y), h1, h2.eq_s(x, y)),
               IsGT(_, _, h1, h2) => IsGT(S(x), S(y), h1, h2.le_succ(S(y), x)) } } }
def (x: Nat).z_le: LE(Z, x) {
   Z => LERefl(Z),
   S(x) => LESucc(Z, x, x.z_le) }
def LE(x, y).le_succ(x y: Nat): LE(S(x), S(y)) {
   LERefl(_) => LERefl(S(x)),
   LESucc(x, y, h) => LESucc(S(x), S(y), h.le_succ(x, y)) }
def LE(S(x), S(y)).le_unsucc(x y: Nat): LE(x, y) {
   LERefl(_) => LERefl(x),
   LESucc(_, _, h) => h.s_le(x, y) }
def LE(S(x), y).s_le(x y: Nat): LE(x, y) {
   LERefl(_) => LESucc(x, x, LERefl(x)),
   LESucc(_, y', h) => LESucc(x, y', h.s_le(x, y')) }
def LE(S(x), y).s_pred(x y: Nat): Eq(Nat, S(y.pred), y) {
   LERefl(_) => Refl(Nat, y),
   LESucc(_, y', _) => Refl(Nat, y) }
```